\documentclass[pageno]{jpaper}

\usepackage{graphicx} 
\usepackage{booktabs} 
\usepackage{tabularx} 
\usepackage{multirow} 
\usepackage{adjustbox} 
\usepackage{float} 
\usepackage{stfloats} 
\usepackage{caption} 
\usepackage{mathptmx} 
\usepackage{amsmath}
\usepackage{fancyhdr}
\usepackage[bottom]{footmisc} 
\usepackage{lipsum} 
\usepackage{tikz} 
\usepackage{soul} 

\usepackage{xcolor} 
\usepackage{enumitem} 
\setlist{nolistsep}
\usepackage[normalem]{ulem}
\usepackage{hyperref} 
\hypersetup{colorlinks,allcolors=black}
\usepackage{amssymb}
\usepackage{pifont}
\usepackage{ifthen} 
\usepackage{xspace} 
\usepackage{comment} 
\usepackage{listings} 
\usepackage[ruled,vlined]{algorithm2e}
\usepackage{subcaption}
\usepackage{makecell}
\usepackage{authblk}

\definecolor{cadmiumgreen}{rgb}{0.0, 0.42, 0.24}
\newcommand{\cmark}{\textcolor{cadmiumgreen}{\ding{51}}}
\definecolor{carnelian}{rgb}{0.7, 0.11, 0.11}
\newcommand{\xmark}{\textcolor{carnelian}{\ding{55}}}

\newcolumntype{L}[1]{>{\raggedright\let\newline\\\arraybackslash\hspace{0pt}}m{#1}}
\newcolumntype{R}[1]{>{\raggedleft\let\newline\\\arraybackslash\hspace{0pt}}m{#1}}

\newcommand{\tdmem}{\textsc{TDMem}\xspace}

\begin{document}
\title{Hardware-assisted Trusted Memory Disaggregation for Secure Far Memory}

\author{Taekyung Heo \qquad Seunghyo Kang \qquad Sanghyeon Lee \qquad Soojin Hwang \qquad Jaehyuk Huh\\School of Computing, KAIST}

\date{}
\maketitle

\thispagestyle{empty}

\begin{abstract}
\label{sec:tdmem-abstract}
Memory disaggregation provides efficient memory utilization across network-connected
systems. It allows a node to use part of memory in remote nodes in the same cluster.
Recent studies have improved RDMA-based memory disaggregation systems, supporting
lower latency and higher bandwidth than the prior generation of disaggregated memory.
However, the current disaggregated memory systems manage remote memory only at coarse
granularity due to the limitation of the access validation mechanism of RDMA. In
such systems, to support fine-grained remote page allocation, the trustworthiness
of all participating systems needs to be assumed, and thus a security breach in a
node can propagate to the entire cluster. From the security perspective, the memory-providing
node must protect its memory from memory-requesting nodes. On the other hand, the
memory-requesting node requires the confidentiality and integrity protection of its
memory contents even if they are stored in remote nodes. To address the weak isolation
support in the current system, this study proposes a novel hardware-assisted memory
disaggregation system. Based on the security features of FPGA, the logic in each
per-node FPGA board provides a secure memory disaggregation engine. With its own
networks, a set of FPGA-based engines form a trusted memory disaggregation system,
which is isolated from the privileged software of each participating node. The secure
memory disaggregation system allows fine-grained memory management in memory-providing
nodes, while the access validation is guaranteed with the hardware-hardened mechanism.
In addition, the proposed system hides the memory access patterns observable from
remote nodes, supporting obliviousness. Our evaluation with FPGA implementation shows
that such fine-grained secure disaggregated memory is feasible with comparable performance
to the latest software-based techniques.
\end{abstract}

\section{Introduction}
\label{sec:tdmem-introduction}
Memory disaggregation has emerged to enable efficient utilization of memory capacity
across system boundaries. The imbalance in memory utilization among virtual machines
or user applications in a cluster necessitates memory capacity sharing among the
nodes in the same cluster. Recent studies showed that RDMA-supporting networks can
allow effective expansion of memory beyond a single node with the low latency and
high bandwidth data transfers~\cite{gu2017efficient, amaro2020can, ruan2020aifm}.
Furthermore, there have been several commercial proposals for new interconnects that
support fine-grained direct memory accesses to remote memory nodes~\cite{cxl, opencapi,
ccix}. Such memory disaggregation reduces the total cost of ownership of data centers
by avoiding over-provisioning of memory capacity for each node.

Although there have been recent studies to improve the performance of the disaggregated
memory systems~\cite{amaro2020can, ruan2020aifm, calciurethinking}, its security
aspect has not been thoroughly investigated. Unlike typical system designs based
on localized memory, the disaggregated memory system opens the memory boundary beyond
the conventional system limit, and thus the memory contents of a user application
can exist across multiple nodes. In addition, a node must allow other nodes to access
part of its memory, being forced to trust the behavior of other nodes. The current
disaggregated memory systems allow only coarse-grained access controls of RDMA, which
limit the flexibility of memory management. To allow fine-grained memory allocation,
a large memory pool must be accessible among nodes, and thus the collective trustworthiness
of all participating nodes must be assumed. In such cases, a security breach in a
single node can potentially propagate to the rest of the nodes since their memories
are shared.

To protect against such \textit{vulnerability propagation}, the disaggregated memory
poses several new challenges. First, the confidentiality and integrity of memory
pages stored in remote nodes must be protected under vulnerable privileged software
in the remote nodes. Second, when a node (\textit{donor node}) donates part of memory
for other nodes, its memory must be protected from any malicious attempt to access
illegitimate regions of memory. Third, memory management must be fine-grained and
flexible. Assigning one big contiguous chunk of memory for sharing leads to the inefficient
utilization of memory capacity, which offsets the benefit of memory sharing for improving
memory utilization. Finally, memory access patterns of memory user nodes must be
indistinguishable from the donor node.

However, the current disaggregated memory studies do not fully address the aforementioned
challenges. A common approach for disaggregated memory is to rely on RDMA supports
in network interface cards for low latency accesses across networks. In such systems,
the remote memory address is stored in the memory user node (\textit{donee node})
for low latency RDMA accesses. On the donor side, RDMA memory regions can be allocated
in one or a few big chunks of memory pool to rely on coarse-grained access validation
of RDMA network devices. Secure fine-grained management of huge memory capacity is
hard to be supported with the limited region-based access control.

\begin{figure}[t!]
    \centering
    \includegraphics[width=0.7\columnwidth]{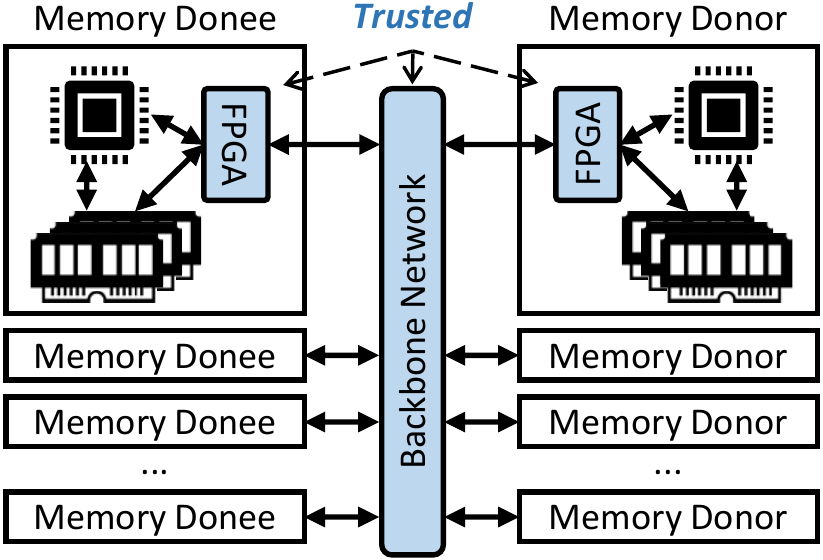}
    \\
    \caption{FPGA-based trusted disaggregated memory}
    \label{fig:fpga-based-trusted-disaggregated-memory}
\end{figure}

To harden the disaggregated memory system, this paper proposes a novel hardware-assisted
disaggregated memory system. For each node, the trusted logic on the FPGA board provides
the fine-grained access validation between different nodes. Based on the security
supports of FPGA, our per-node memory disaggregation hardware engines form a trusted
memory disaggregation system for secure page transfers. As the confidentiality and
integrity of the FPGA logic and its states are guaranteed, the operations on the
FPGA board are completely protected from vulnerable operating systems. In addition,
the FPGA boards are connected with their own networks, which do not pass through
the network stack of the vulnerable operating systems. The new hardware-assisted
memory disaggregation system provides strong isolation among the participating nodes,
and thus even if one of the nodes is attacked, it does not adversely affect the other
nodes. Figure~\ref{fig:fpga-based-trusted-disaggregated-memory} presents the trusted
FPGA-based disaggregated memory system.

To overcome the limitation of the current coarse-grained permission validation, the
hardware engine provides a fine-grained page-granular access control. To allow low
latency access, the OS of the donee node maintains the target page address in the
remote node. However, the secure FPGA logic in the donor node checks the source of
page requests and validates the access at page granularity. Such decoupling of address
translation and permission validation allows fine-grained access validation without
degrading the performance.

Another important aspect of security is the patterns of page accesses. Prior work
showed that such coarse-grained fault address patterns can leak critical information~\cite{xu2015controlled}.
As memory disaggregation systems such as CXL have fixed mappings between the donee's
address space and the donor's address space, a donor can track the memory access
patterns of donees. To hide the memory access patterns, our proposal relies on the
swap subsystem of the Linux kernel, which allocates a new swap slot on every page
swap. As a result, the mappings between two address spaces constantly change, breaking
up the connection between two address spaces. In addition, with the assistance of
FPGAs, \tdmem further obfuscates write memory access patterns.

We propose Trusted Disaggregated Memory, \tdmem, the hardware design for secure and
fine-grained disaggregated memory. This new hardware-assisted system is implemented
in a Linux system equipped with the Xilinx Alveo U50. The Linux kernel has been modified
for the tight integration of \tdmem with the existing virtual memory system. Our
evaluation shows that even with the security supports with fine-grained memory management,
the performance degradation is minimized to 4.4\%, compared to the prior best-performing
RDMA-based disaggregated memory system.

The source codes for FPGA logic and Linux patch will become publicly available after
publication. The main contributions of this paper are as follows:

\begin{itemize}
    \item This paper discusses the potential security problem of disaggregated memory
    and proposes a hardware-assisted approach by using the secure FPGA technique.
    \item It overcomes the limitation of the current RDMA-based approach with coarse-grained
    access control and provides the secure fine-grained page-level allocation.
    \item The hardware-assisted mechanism allows fast page transfers comparable
    to the prior mechanism based on region allocation.
    \item It obfuscates the memory access pattern of donees by exploiting the existing
    swap subsystem, so the data leaks by page access patterns are avoided.
\end{itemize}

The rest of the paper is organized as follows. Section 2 presents the background
and motivation for hardware-assisted secure memory disaggregation. Section 3 discusses
the threat model and security challenges. Section 4 presents the design of \tdmem,
and Section 5 describes its implementation. Section 6 presents the experimental results
with the real FPGA implementation. Section 7 discusses the related work, and Section
8 concludes the paper.

\section{Background and Motivation}
\label{sec:tdmem-background-and-motivation}
\subsection{Memory Disaggregation}
Memory disaggregation allows data to reside in remote memory. In general, a disaggregated
memory system has several nodes connected with a high-speed network~\cite{gu2017efficient,
amaro2020can, ruan2020aifm}. In this paper, we denote a node that donates its memory
as a \textit{donor} and a node that utilizes the memory of other nodes as a \textit{donee}.
A node can be a donor and a donee at the same time. From the perspective of a memory-consuming
application, there are two types of memories: local memory and remote memory. The
local memory is the memory in a donee that can be accessed without additional network
latency and address translations. The remote memory is the memory donated by a donor.

Remote Direct Memory Access (RDMA), which enables direct memory access to the memory
in a remote node, has been widely used in prior memory disaggregation studies~\cite{gu2017efficient,
amaro2020can, ruan2020aifm}. The communication channel of RDMA is called a queue
pair. A queue pair is composed of a send queue and receive queue. The communication
between hosts can be done with the abstraction called \textit{verbs}. Verbs are classified
into two types depending on the involvement of a remote CPU: one-sided verbs and
two-sided verbs. One-sided verbs do not require a remote CPU's involvement, while
two-sided verbs require. Therefore, one-sided verbs are more scalable than the other.
However, one-sided verbs are not a free lunch. A memory region must be registered
to access memory with one-sided verbs. The memory registration takes tens of microseconds~\cite{frey2009rdmacost}.
Accesses from remote nodes to a memory region are protected with a key, which is
called \texttt{rkey}.

\subsection{Requirements for Secure Memory Disaggregation}
The distributed and shared nature of a disaggregated memory system imposes several
requirements. In this subsection, we discuss four specific requirements that have
been less explored in this field: memory protection, secure and fine-grained memory
allocation, fast and secure address translation, and memory access pattern obfuscation.

\newcolumntype{A}{>{\centering\arraybackslash}m{7.5em}}
\newcolumntype{B}{>{\centering\arraybackslash}m{6.5em}}
\begin{table*}[t!]
    \footnotesize
    \centering
    \begin{tabular}{A|B|B|B|B|B|B}
    \toprule
        Name &
            Memory Protection &
            \hspace{0.7em}Secure\newline Memory Alloc. &
            Fine-grained\newline Memory Alloc.&
            Address Translation &
            Secure Translation &
            Memory Access Obfuscation\\
    \midrule
        Intel CXL~\cite{cxl} &
            \cmark &
            \cmark &
            \xmark &
            Direct &
            \cmark &
            \xmark \\
        Kona~\cite{calciurethinking} &
            \xmark &
            \xmark &
            \xmark &
            Indirect &
            \xmark &
            \xmark \\
        DeACT~\cite{kommareddy2020deact} &
            \xmark &
            \xmark &
            \xmark &
            Indirect &
            \xmark &
            \xmark \\
        ThymesisFlow~\cite{pinto2020thymesisflow} &
            \xmark &
            \xmark &
            \xmark &
            Direct &
            \xmark &
            \xmark \\
        AIFM~\cite{ruan2020aifm} &
            \xmark &
            \xmark &
            \cmark &
            Indirect &
            \xmark &
            \xmark \\
        Fastswap~\cite{amaro2020can} &
            \xmark &
            \xmark &
            \xmark &
            Indirect &
            \xmark &
            \cmark \\
        InfiniSwap~\cite{gu2017efficient} &
            \xmark &
            \xmark &
            \xmark &
            Indirect &
            \xmark &
            \cmark \\
    \midrule
        \tdmem &
            \cmark &
            \cmark &
            \cmark &
            Indirect &
            \cmark &
            \cmark \\
    \bottomrule
    \end{tabular}
    \\
    \caption{Comparison with prior studies}
    \label{tab:prior-work-comparison-table}
\end{table*}

\noindent
\textbf{Memory protection:}
The confidentiality and integrity of pages stored in remote memory must be guaranteed.
As a donor has full control over donated memory, a donor may read or write any stored
pages. Moreover, if a disaggregated memory system fails to isolate nodes due to design
flaws or bugs, a malicious donee may read or write unauthorized pages. Therefore,
pages should be encrypted before being swapped out to remote memory. In addition,
any modifications to pages have to be detected to prevent reading contaminated pages.
Reading contaminated pages may result in a malfunction of applications or privilege
escalation.

\noindent
\textbf{Secure and fine-grained memory allocation:}
First, memory allocation metadata must be securely protected from adversaries. The
metadata include the start address, size, and ownership of allocated memory. As these
metadata are used for memory access control, tampering of memory allocation metadata
may result in unauthorized reads or writes of memory. Second, memory allocation must
be done at fine-granularity for efficient memory utilization. Coarse-granular memory
allocation may fail to meet the fluctuating memory demands of applications, which
results in wasted memory from internal fragmentation. Fine-granular memory allocation
reduces the wasted memory from internal fragmentation.

\noindent
\textbf{Fast and secure address translation:}
Memory disaggregation involves mappings between a local address space and a remote
address space. The local address space is used by a donee's CPUs to access data in
local memory, and part of the local address space can be backed by the remote memory
in donors. As the address translations from a local address to a remote address are
in the critical path of accessing data, fast address translation is essential to
achieve high performance. In addition to fast identification of remote addresses,
secure address translation is critical to prevent a compromised donor from providing
corrupted pages and to prohibit a malicious donee from reading unauthorized pages.

\noindent
\textbf{Memory access pattern obfuscation:}
The memory access patterns of donees must be hidden from a donor node. Even if a
donor node can observe the memory access patterns, the patterns should not convey
any critical information by obfuscating the memory access pattern. Many prior studies
have shown that the disclosure of memory access patterns may leak critical information
such as the access frequency of data and correlation between them~\cite{zhuang2004hide,
islam2012access, maas2013phantom}. Moreover, several practical software attack schemes
based on memory access pattern detection were proposed~\cite{osvik2006aesattack,
islam2012access}. A malicious observer can extract information about the location
of private data from memory access patterns, which can result in the revelation of
private data.

\subsection{Prior Studies in Memory Disaggregation}
This subsection revisits and compares prior memory disaggregation studies to find
whether the aforementioned requirements are met. Table~\ref{tab:prior-work-comparison-table}
summarizes the prior studies with respect to the aforementioned requirements.

\noindent
\textbf{Memory protection:}
CXL offers the confidentiality, integrity, and replay protection of data. However,
other prior studies support none of them. RDMA-based disaggregated memory systems~\cite{ruan2020aifm,
amaro2020can, gu2017efficient} rely on the default memory protection mechanism, which
is to authorize accesses to a memory region with \texttt{rkey}. However, as prior
studies have shown~\cite{simpson2020securing, rothenberger2021redmark}, \texttt{rkeys}
are predictable because of the small key size and unrobust implementations. These
security holes let a malicious donee read or write an unauthorized memory region.

\noindent
\textbf{Secure and fine-grained memory allocation:}
In CXL, memory allocation is done at the granularity of logical devices. CXL defines
a memory pool device as a memory device that can be shared by multiple donees. A
memory pool device can have 16 logical devices at maximum. The allocation of logical
devices is advertised through a designated memory area, which is called \texttt{DVSEC}.
As \texttt{DVSEC} can be configured by a fabric manager (FM) only, as long as the
FM is securely protected, memory allocation metadata are securely protected. Except
for CXL, prior studies do not describe the security model for secure memory allocation.
In short, the memory allocation of CXL is secure but done at coarse-granularity.

The security feature for memory allocation is not clarified in the other studies.
Moreover, most prior studies allocate memory at coarse-granularity in the initialization
stage. ThymesisFlow allocates memory at the unit of a section, which is the memory
management unit in the sparse memory model of the Linux kernel. The default section
size is set to 1TB according to its specification. Kona allocates memory using a
coarse-granular slab. Fastswap reserves memory in a donor where the memory size matches
the size of the swap device size of the donee. The reserved memory in the donor cannot
be shared with other donees. Unlike the other studies, AIFM allocates memory at object-granularity
on demand. However, it loses transparency, mandating modifications to the user applications.

\noindent
\textbf{Fast and secure address translation:}
Prior studies can be classified into two types according to the existence of additional
address space between a local address space and a remote address space. The studies
without additional address space are classified as \textit{direct}, and others are
classified as \textit{indirect}. Direct has an advantage in terms of performance
because it can avoid additional address translations. However, it may weaken the
security model by exposing the physical address space of a donor to donees directly.
Moreover, donees may suffer from the leak of memory access patterns as the mappings
between the local addresses and remote addresses are fixed. CXL and ThymesisFlow
adopt direct translations. Kona is classified as indirect because it virtualizes
remote memory with the support of FPGAs. While swap-based memory disaggregation does
not have a global shared address space like Kona, it has an additional address space
because of the Linux kernel design. On page swaps, a local physical address has
to be translated to a swap offset first in the swap address space. The corresponding
remote physical address is accessed with the swap offset.

Secure address translations are not supported or undefined in most prior studies
except for CXL. As CXL keeps logical device allocation metadata in \texttt{DVSEC},
which cannot be accessed by non-FM, address translation is securely protected from
adversaries. However, other prior studies do not describe any mechanism to make address
translation secure. The prior studies rather believe that participating nodes are
trustworthy and do not violate the protocol.

\noindent
\textbf{Memory access pattern obfuscation:}
Memory disaggregation systems with direct address translations have a critical limitation
that memory access patterns may leak. As the mappings between the donee's address
space and the donor's address space are fixed, a malicious donor may track the memory
access pattern of a donee by tracking the access pattern at the donor side. One of
the solutions to hide memory access patterns is the Oblivious RAM (ORAM), which is
an algorithm that obfuscates memory access patterns from adversaries. Although there
have been many studies to improve the performance of ORAM and to scale them~~\cite{yu2015proram,
ren2015constants, chan2017circuit, doerner2017scaling, wang2017cooperative}, the
adoption of ORAM still incurs prohibitive performance degradation in the real world.
As an alternative, the existing swap system in the Linux kernel already has a sort
of memory access pattern obfuscation mechanism. As the mappings between the local
address space and the swap address space change on every update of pages, there is
a weak relationship between the local memory access patterns and the remote memory
access patterns.

\subsection{FPGA}
Field-programmable Gate Array (FPGA) is a device that allows the programming of on-chip
hardware logic resources after fabrication. Programming of an FPGA can be done with
a bitstream, which contains a sequence of commands. Modern FPGA boards support security
and network features.

\noindent
\textbf{Security supports:}
As this study has been conducted with Xilinx FPGA boards, this subsection introduces
the security features of Xilinx FPGAs. Xilinx FPGAs offer several security features~\cite{peterson2013developing}.
First, bitstream encryption is a feature that encrypts a bitstream with a key. Designing
an accelerator is a costly task. Therefore, there is a need to prevent unauthorized
users from deploying a proprietary bitstream. When a bitstream is encrypted, the
bitstream can be programmed only when a user has a valid decryption key. Second,
the bitstream authentication validates a bitstream. A bitstream is authenticated
with a keyed-hashed message authentication code (HMAC). Modifications to a bitstream
can be detected with the HMAC. Third, disabling a JTAG port prevents tampering attempts
to an FPGA. The bitstream encryption and authentication do not prevent an FPGA from
being reprogrammed with an arbitrary bitstream. Reprogramming of an FPGA can be permanently
blocked by disabling a JTAG port. A JTAG port can be disabled by blowing up a corresponding
electronic FUSE (eFUSE).

\noindent
\textbf{Separate high-speed network:}
Modern FPGA boards are equipped with high-speed network modules. An FPGA board from
Xilinx, Alveo U50, has a QSFP28 network module. The network module can be used for
any network protocol as long as the protocol can be implemented in the FPGA. A network
protocol can be implemented with the IP cores offered by Xilinx~\cite{cmac_usplus}
or open-source projects~\cite{sidler2015scalable, ruiz2019limago, forencich2020corundum}.
Note that the network between FPGAs can form a physically separate network from the
host's. In other words, the network stack of FPGAs does not rely on the operating
system. The direct network connection between FPGAs enables a bump-in-the-wire acceleration
of incoming packets at high speed.

\section{Threat Model}
\label{sec:tdmem-threat-model}
\begin{figure}[t!]
    \vspace{-1em}
    \hspace{1cm}
    \centering
    \includegraphics[width=\columnwidth]{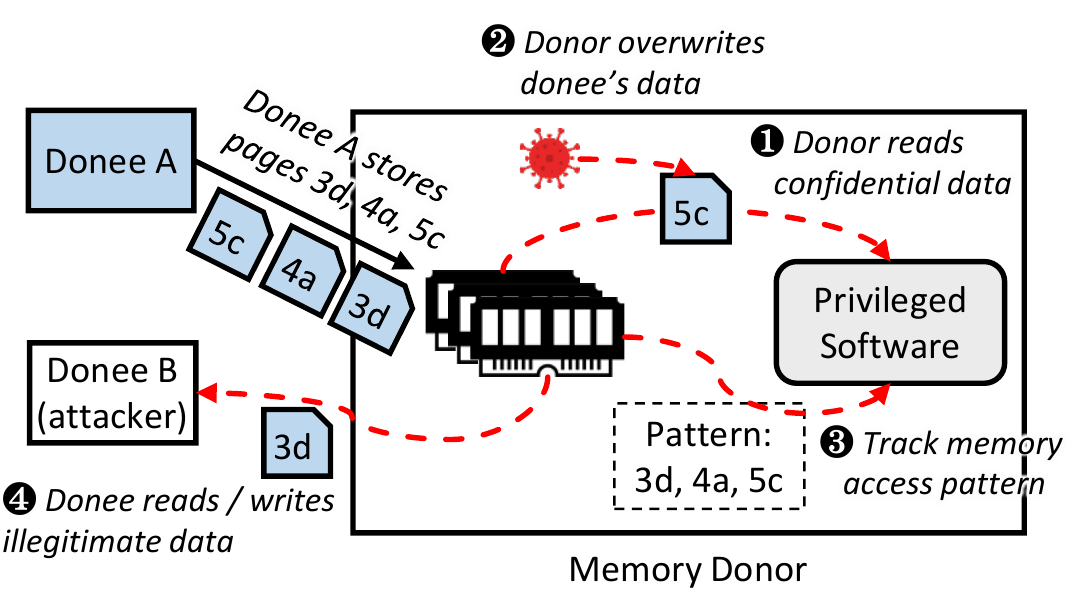}
    \\
    \caption{\tdmem threat model}
    \label{fig:threat-model}
\end{figure}

This study assumes a threat model where all participating nodes mutually distrust
each other since our goal is to prevent the propagation of attacks across the nodes.
We assume that a donor or donee can be malicious or compromised at any time. A cloud
service operator may attempt to steal the precious intellectual property of users.
Alternatively, a cloud user may subvert the virtualization layer and take the root
privilege. Initially, the attacker may have a limited privilege and want to gain
more privileges of other nodes. Figure~\ref{fig:threat-model} illustrates the threat
model of \tdmem. As a donor has full control of donated memory, a donor may read
donees' confidential pages or overwrite them. Moreover, a compromised donor may attempt
to observe which pages are written. It is not possible to identify which pages are
accessed by DMA reads. However, DMA writes can be monitored indirectly by the OS.
The OS can periodically read the contents of pages from the donees, and check any
changes of contents. With such monitoring, the OS can identify which pages are accessed.
Donees cannot be trusted either. A malicious donee may try to violate the protocol
and try to read or write unauthorized pages.

All nodes are equipped with an FPGA board that implements the secure memory disaggregation
engine. The bitstream is encrypted with a secret key, preventing an unauthorized
node from participating in the memory disaggregation network. The authenticity of
a bitstream is guaranteed by the bitstream authentication mechanism. The JTAG ports
of FPGA boards are disabled after programming the bitstream. FPGA boards are assumed
to be securely protected from the operating system. Once the FPGA logic is loaded,
its operation is completely isolated from the operating system. The separate network
path of an FPGA board allows complete isolation from the operating system. However,
since the interconnects and memory are physically exposed, we do not consider physical
attacks on donor or donee nodes. The availability is not guaranteed either since
a malicious OS can block the operation of an FPGA board.

The trusted computing base (TCB) is defined differently by the role of a node. A
donee node trusts its kernel and all FPGA boards in the system. A donee relies on
the swap subsystem of the Linux kernel. Therefore, if its kernel is compromised and
the kernel reads or modifies pages on the swap path, the attack cannot be eschewed.
Considering that the goal of \tdmem is to prevent the propagation of attacks across
nodes, the definition of the TCB is reasonable. On the other hand, a donor node trusts
all FPGA boards. It does not have to trust its or others' kernel.

\section{Design}
\label{sec:tdmem-design}
\subsection{Overview}
\tdmem is a disaggregated memory system with hardware supports to enhance the security
of disaggregated memory. \tdmem has multiple participating nodes in a system, and
each node is equipped with an FPGA board. A node can be a donor or/and donee. Workloads
running in a donee can utilize the memory of remote nodes when needed.

\vspace{0.5em}
The followings are the design goals of \tdmem.
\vspace{0.2em}
\begin{enumerate}
    \item The confidentiality of donees' pages is guaranteed and their integrity is validated.
        \vspace{0.1em}
    \item Memory allocation is done at 4KB page granularity and memory allocation metadata
        is protected from malicious OSes.
        \vspace{0.1em}
    \item Address translations are securely protected and cannot be bypassed or forged.
        \vspace{0.1em}
    \item The memory access patterns of donees should be indistinguishable from the donor OS.
\end{enumerate}
\vspace{0.5em}

\begin{figure}[t!]
    \hspace{-1cm}
    \centering
    \includegraphics[width=0.95\columnwidth]{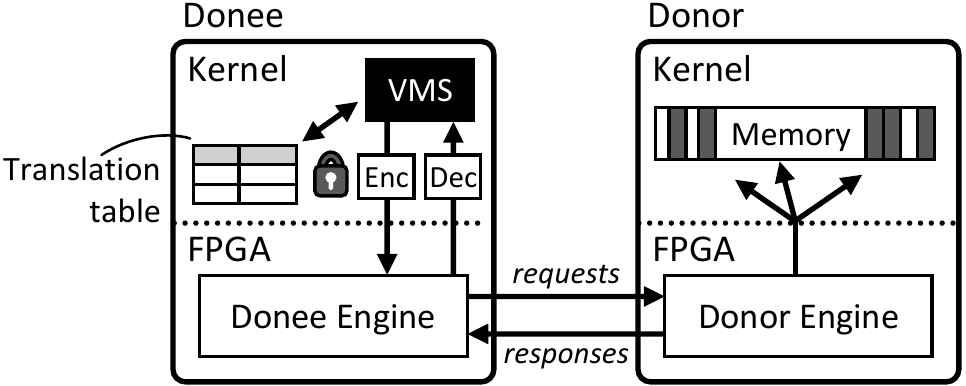}
    \\
    \caption{Overview of \tdmem}
    \label{fig:tdmem-design-overview}
\end{figure}

Figure~\ref{fig:tdmem-design-overview} presents the overview of \tdmem. The donee's
virtual memory system (VMS) is tightly integrated to the secure memory disaggregation
engine using the frontswap interface. Frontswap allows the Linux kernel to redirect
page swap requests to other subsystems. In \tdmem, the requests are redirected to
the donee engine. Swap-out and swap-in commands from the donee are forwarded to its
FPGA. The commands are parsed and sent to the donor engine, which is responsible
for memory allocation and permission checks. On page faults, swapped-out pages are
read from the donor, passing through the secure memory disaggregation engines. In
the rest of this section, how the design goals are met will be described.

\subsection{Confidentiality and Integrity Validation}
\tdmem guarantees the confidentiality of donees' pages and validates their integrity
with the donee-side software page encryption. When a page is evicted from a donee
node, the page is encrypted with AES-GCM in the donee-side kernel. The encryption
key is randomly generated during the initialization of \tdmem. The key resides in the
kernel and is tied to a node, not shared with others. As encrypted pages cannot be
read without keys, the confidentiality of pages is guaranteed. The integrity of pages
can be validated by encryption also. As AES-GCM supports the authentication of encrypted
data, any modifications to encrypted pages can be detected on decryption. The authentication
can be done with a media authentication code (MAC). On page encryption and decryption,
a MAC is generated that is unique to the page content. Therefore, once an encrypted
page is modified, the MAC generated on decryption does not match to the previous
one generated on encryption. MACs are stored in the donee-side address translation
table.

\begin{table}[t!]
    \footnotesize
    \centering
    \begin{tabular}{c|c|c|c}
        \toprule
        \multicolumn{2}{c|}{\textbf{CPU}}                         & \multicolumn{2}{c}{\textbf{FPGA}}                      \\ \toprule
        \textbf{Encryption}          & \textbf{Decryption}         & \textbf{Encryption}        & \textbf{Decryption}        \\ \hline
        \multicolumn{1}{r|}{2.07us} & \multicolumn{1}{r|}{2.09us} & \multicolumn{1}{r|}{152.58us} & \multicolumn{1}{r}{154.62us} \\ \hline
    \end{tabular}
    \caption{AES-GCM latency comparison}
    \label{tab:aes-gcm-latency-comparison}
\end{table}

We decide to encrypt pages in the kernel not in the FPGA because of two reasons.
First, in terms of the performance and cost of memory encryption, CPUs are superior
to FPGAs. Intel CPUs already have highly optimized encryption engines, which is called
AES-NI~\cite{akdemir2010breakthrough}. AES-NI is highly optimized and operating at
a very high frequency (several GHz) compared to FPGAs (200-250MHz). Table~\ref{tab:aes-gcm-latency-comparison}
compares the latencies of CPUs and FPGAs on encryption and decryption of a 4KB page.
The CPU latency is measured with the \texttt{tcrypt} module in the Linux kernel.
The FPGA latency is estimated with the Vitis HLS tool, where the operating frequency
is set to 250MHz. We use an open-source AES-GCM implementation provided by Xilinx~\cite{vitislibraries}.
On CPUs, it takes 2.07us and 2.09us to encrypt and decrypt a 4KB page, respectively.
On the other hand, on FPGAs, it takes 152.58us and 154.62us for encryption and decryption,
respectively. Moreover, the resource utilization of encryption engines is also a
problem. An AES-GCM encryption engine consumes 5.7\% of LUT and 4.7\% of FF. A decryption
engine takes 5.9\% of LUT and 5.0\% of FF. Considering that the performance of the
FPGA-based encryption engine is much lower than CPUs, the resource utilization is
not reasonable.

\subsection{Fine-grained and Secure Memory Allocation}
\tdmem supports fine-grained memory allocation with the assistance of FPGAs. \tdmem
has memory allocators in the secure memory disaggregation engines. In prior RDMA-based
disaggregated memory, donated memory is allocated and registered at coarse-granularity
in the initialization stage. Once a memory region is registered by a donee node,
the allocated memory can be accessed by the node and cannot be shared with others.
In such an RDMA-based system, registering memory at fine-granularity frequently is
not a reasonable solution because of the excessive overheads of registering new regions~\cite{frey2009rdmacost}.

An alternative way to provide secure fine-grained allocation is to use two-sided
verbs of RDMA or common TCP/IP network stacks. In that approach, the disaggregation
software component in donor nodes will receive page access requests through networks
and validate them individually. However, it can significantly increase the latency
and limit the bandwidth since the software layer on the donor side must handle that.
Requests and responses must be sent to the software layer by the long network stack
through the NIC and operating system.

To overcome the limitation, disaggregated memory must provide an efficient page-level
access control. The memory capacity of a donor node must be dynamically allocated
to different donees or its own applications for efficient memory utilization. The
page allocator in the donor must be nimble, and the hardware permission validation
must be designed to support such page-level validation. \tdmem has memory allocators
in FPGAs to allocate memory at 4KB page granularity on every page swap-out request.
Memory allocators are located in FPGAs to minimize the latency on memory allocation.
With FPGA-based implementation, memory allocation requests can be handled immediately
once a command reaches the FPGA. The memory allocators are securely protected from
malicious OSes because both the logic and metadata reside in FPGAs and cannot be
altered by OSes.

\subsection{Decoupled Translation and Permission Checks}
\begin{figure}[t!]
    \centering
    \includegraphics[width=0.93\columnwidth]{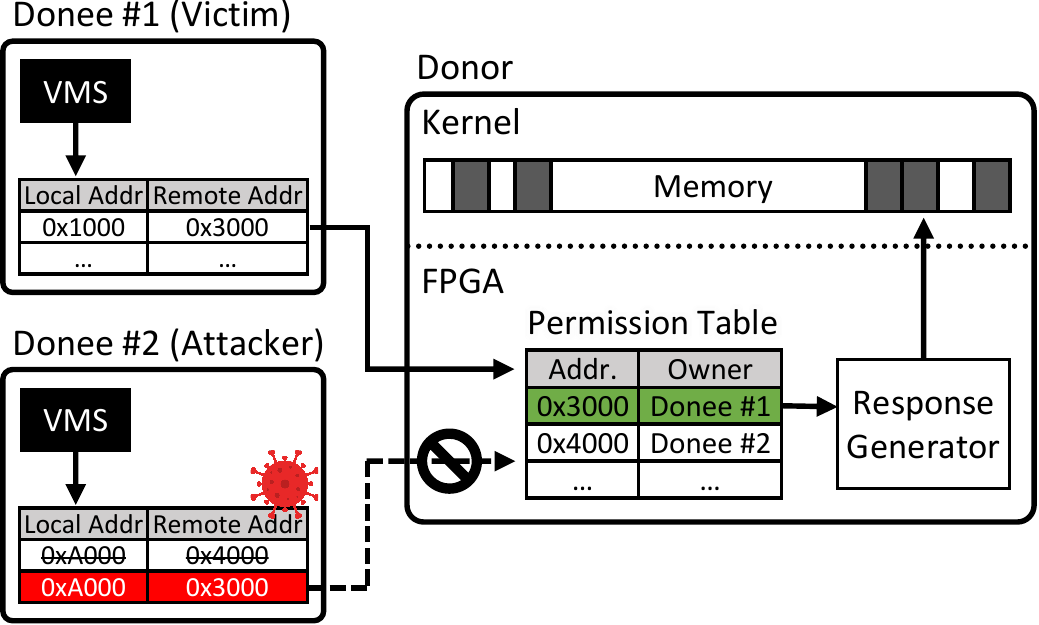}
    \\
    \caption{Decoupled translation and permission checks}
    \label{fig:decoupled-translation-and-permission-checks}
\end{figure}

Address translation must be high-performant and secure because it is in the critical
path and controls memory accesses. \tdmem achieves both goals by decoupling address
translations and permission checks. Figure~\ref{fig:decoupled-translation-and-permission-checks}
illustrates the decoupled address translation and permission checks. To achieve
high-performance, \tdmem avoids introducing an additional address space. \tdmem exposes
the donor's physical address space to donees directly and offloads address translations
to donees. Each donee has a mapping table that maps swap offsets to remote physical
addresses. By exposing the remote address space directly to donees, performance degradation
is avoided. As address translations are done in donees, a compromised donor cannot
forge the translation table.

The key problem with offloading address translations to donees is the risk of donees'
accesses to unauthorized pages. A malicious donee may try to read illegitimate pages
by tampering the translation table. \tdmem tackles the problem with a permission
table at the donor engine. The table tracks the ownership for all memory pages in
the donor. On page loads, the permission table is looked up to validate the load
request. The permission table is updated on memory allocation and deallocation. Another
problem with having a permission table at the donor side is that a compromised donor
may manipulate the permission table. \tdmem blocks this type of attack by having
the permission table in the FPGA board. The FPGA board has a donor engine where the
design is securely loaded and cannot be tampered. The integrity of the donor engine
is guaranteed because the FPGA logic is protected from malicious privileged software.

\subsection{Memory Access Pattern Obfuscation}
\tdmem obfuscates the donees' memory access patterns to a donor. Disaggregated memory
systems with direct address translations have limitations that donees' memory access
patterns can be observed by a malicious donor. \tdmem overcomes this limitation with
the adoption of the swap subsystem of the Linux kernel. In the Linux kernel, the swap
subsystem has a swap address space in addition to the virtual and physical address
spaces. The swap address space is used to allocate swap slots on swap-out requests.
The swap slot (swap offset) is used to determine the location of the swapped-out
page in a swap device. The allocated slot is deallocated on swap-in requests. With
this swap allocation mechanism, a swap device can be fully utilized. As the mappings
between the physical address between the swap address are not fixed, memory access
patterns can be obfuscated.

\begin{figure}[t!]
    \centering
    \includegraphics[width=\columnwidth]{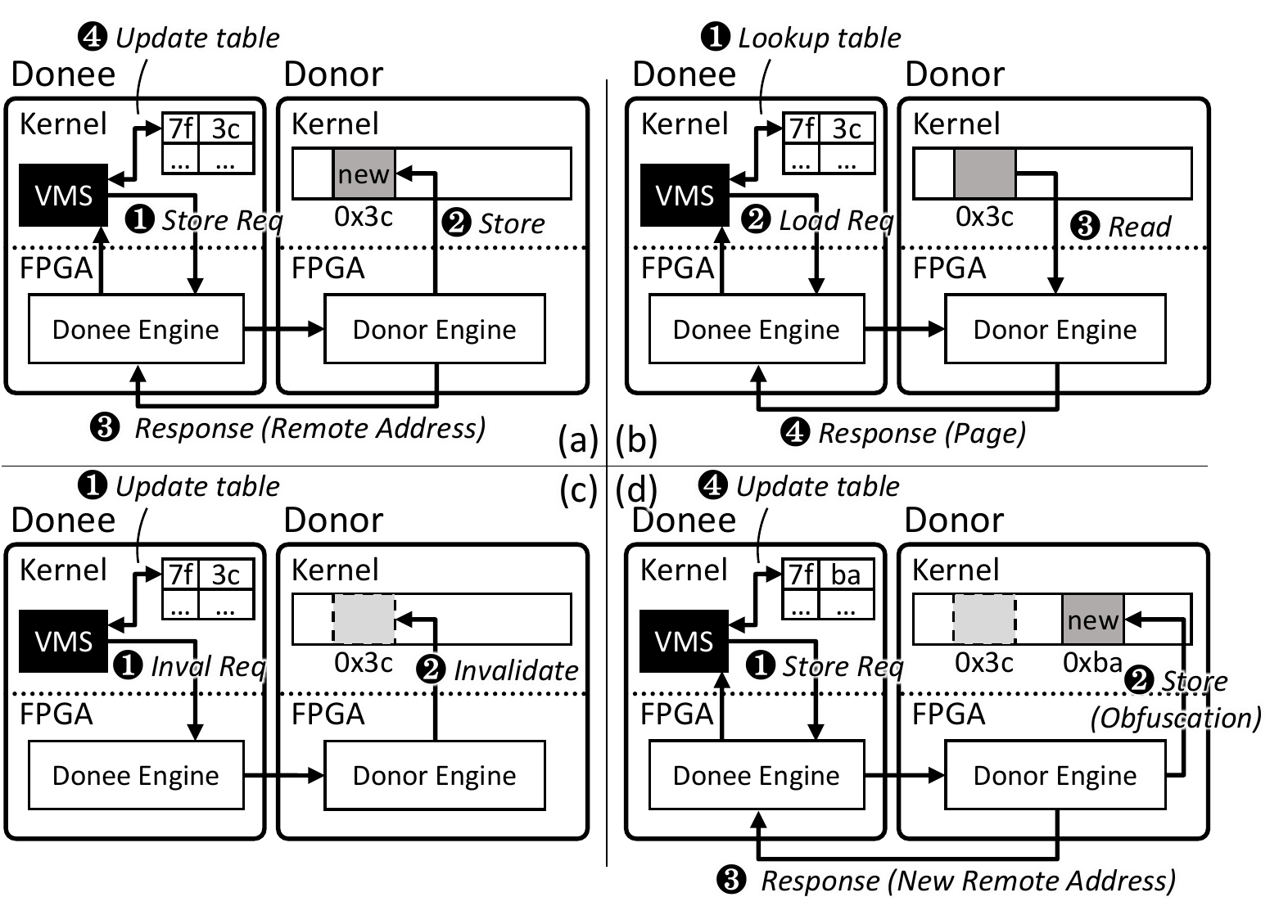}
    \\
    \caption{Memory access pattern obfuscation mechanism.
    Subfigure (a), (b), (c), (d) illustrate store, load, invalidate page, and store, respectively.}
    \label{fig:memory-access-pattern-obfuscation}
\end{figure}

Although the swap slot allocation logic provides a sort of obfuscation, it has a
limitation that the logic is known by an attacker as the Linux kernel is open source.
\tdmem overcomes the limitation by strengthening memory access pattern obfuscation
with the support of FPGAs. \tdmem further obfuscates memory access patterns by changing
the remote address of a local page with the secret logic implemented in FPGAs. To
integrate the logic with the Linux kernel, \tdmem borrows the commands from the frontswap
interface. The frontswap interface has four commands: store, load, invalidate page,
and invalidate area. On swapping out of a page, a store command is issued. On swapping
in of a page, a load command and invalidate page command are generated. A load command
is used to read the page from the donor. An invalidate page command deallocates the
page from the donor. An invalidate area command is issued on the shutdown of a machine,
which deallocates all allocated swap pages. Secret logic allocates a different page
on every store command to enhance memory access pattern obfuscation.

Figure~\ref{fig:memory-access-pattern-obfuscation} illustrates how the frontswap-like
interface can be used for memory access pattern obfuscation. Figure~\ref{fig:memory-access-pattern-obfuscation}
(a) shows the case where a store occurs on a page, which is at 0x7f in the donee
address space. The donor engine then allocates a new page at 0x3c in the remote memory,
which is randomly selected among free pages. The remote address, 0x3c, is sent back
to the donee engine. As shown in Figure~\ref{fig:memory-access-pattern-obfuscation}
(b), on loading of the stored page, the translation table is looked up by VMS. After
that, a load command is forwarded to the donor engine to load the page at 0x3c. When
a page is outdated, the VMS invalidates the stored page in remote memory with invalidation,
as shown in Figure~\ref{fig:memory-access-pattern-obfuscation} (c). Figure~\ref{fig:memory-access-pattern-obfuscation}
(d) shows storing of the same page after invalidation. Although the donee-side page
offset is the same as the previous one, a different free page on remote memory should
be allocated. Such memory access pattern obfuscation prevents a malicious donor from
peeking page access pattern.

\section{Implementation}
\label{sec:tdmem-implementation}
\subsection{Overview}
\label{sec:tdmem-implementation-overview}
\tdmem is implemented in a Linux system equipped with the Xilinx Alveo U50. The FPGA
board has an FPGA chip, network module, and 8GB on-board HBM. The FPGA chip is used
to implement the \tdmem logic, and the network module is used to connect nodes with
100Gbps Ethernet. The role of HBM memory is determined by the role of a node. A donee
uses the whole HBM memory as remote memory that can be accessed without network latency.
On the other hand, a donor uses the HBM memory for donated memory and memory allocation
metadata storage. The reason for this design choice will be described in the rest
of this section. To summarize, \tdmem has three memory tiers: donee-HBM, donor-HBM,
and donor-DRAM. On a page swap out request, a donee can specify the target memory
tier to store the page. IP cores are written in Vitis HLS and Verilog, and they are
integrated in the Vivado flow.

\begin{figure}[t!]
    \centering
    \includegraphics[width=0.8\columnwidth]{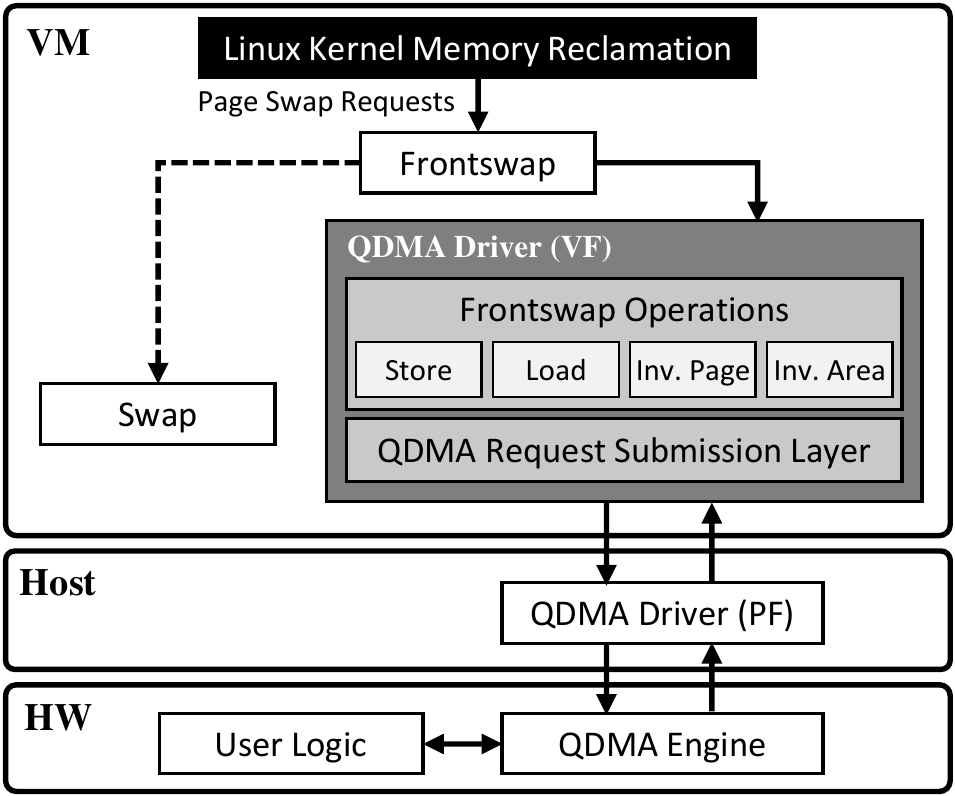}
    \\
    \caption{Software and hardware stacks of a donee}
    \label{fig:donee-sw-hw-stack}
\end{figure}

\subsection{\tdmem Operations}
\label{sec:tdmem-implementation-operations}
\tdmem is tightly integrated to the virtual memory system of the Linux kernel. The
integration is done with frontswap, which is the interface that redirects swap operations
to other subsystems. Frontswap has four functions: store, load, invalidate page,
and invalidate area. These functions become the default operations of \tdmem. On
each operation, a command is generated in the kernel driver and forwarded to the
donee engine over PCIe. Figure~\ref{fig:donee-sw-hw-stack} illustrates the software
and hardware stacks of a donee, illustrating the entire flow. Please note that the
virtualization layer is used to isolate kernel driver bugs from the host machine.

\begin{figure}[t!]
    \centering
    \includegraphics[width=\columnwidth]{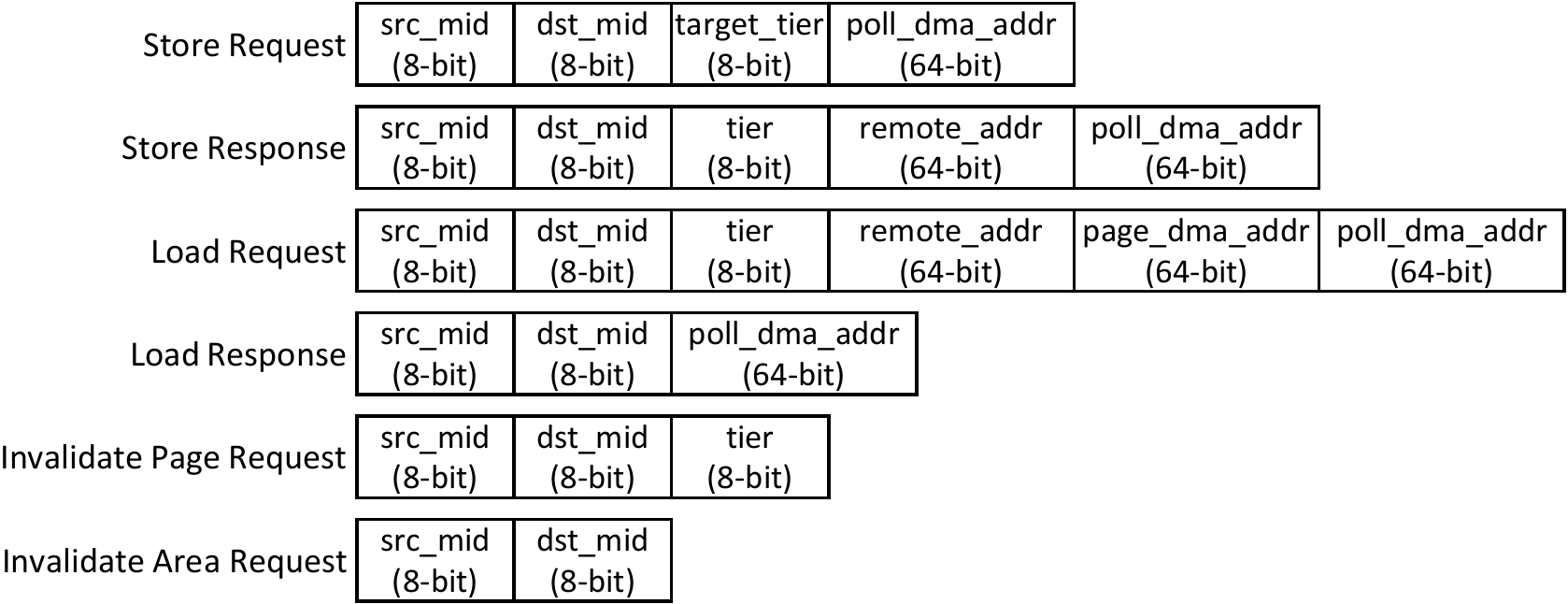}
    \\
    \caption{Command format}
    \label{fig:commands}
\end{figure}

\begin{table}[t!]
    \scriptsize
    \centering
    \setlength\extrarowheight{1pt}
    \begin{tabular}{c|c|L{4cm}}
        \toprule
        \multicolumn{1}{c}{\textbf{Command Field}}
            & \multicolumn{1}{c}{\textbf{Bit-width}}
            & \multicolumn{1}{c}{\textbf{Description}}\\ \toprule
        \texttt{src\_mid} & 8-bit & MID of the source machine \\
        \texttt{dst\_mid} & 8-bit & MID of the destination machine \\
        \texttt{tier} & 8-bit & Memory tier where the page is stored \\
        \texttt{target\_tier} & 8-bit & Target memory tier to store the page \\
        \texttt{remote\_addr} & 64-bit & Memory address in the corresponding memory tier \\
        \texttt{poll\_dma\_addr} & 64-bit & DMA address of a completion \\
        \texttt{page\_dma\_addr} & 64-bit & DMA address of a page to load\\
        \bottomrule
    \end{tabular}
    \caption{Command field descriptions}
    \label{tab:command-field-descriptions}
\end{table}

The commands generated by the kernel are handled by the FPGA engines. The size of
a command is 64B, which is the default transfer size of the DMA engine. A command
encodes required information to process the command. The command fields and their
descriptions are described in Figure~\ref{fig:commands} and Table~\ref{tab:command-field-descriptions}.
The completion of a command is identified by polling. The donee kernel is responsible
for reserving a memory block that is accessible from its FPGA board. The memory block
is named as \texttt{completion}, and a \texttt{completion} contains several metadata
required to process the operation. A \texttt{completion} for a store command contains
the completion status of a request, stored memory tier, and remote memory address.
These metadata are kept in the translation table in the donee. To load the page,
the donee generates a request using the metadata.

\subsection{Hardware Implementation}
\label{sec:tdmem-hardware-implementation}

{
\tolerance=1
\emergencystretch=\maxdimen
\hyphenpenalty=10000
\hbadness=10000

\subsubsection{Common Components}
\tdmem uses the Xilinx QDMA Subsystem for PCI Express~\cite{qdma} for the communication
between the host and card. QDMA is chosen over XDMA because it supports a queue-based
submission mechanism, which supports thousands of concurrent requests. This feature
is essential to serve concurrent page swaps. The descriptor bypass mode has been
enabled to allow the card to write the host DRAM directly. For the networking feature,
the UltraScale+ Integrated 100G Ethernet Subsystem~\cite{cmac_usplus} is used for
the prototyping purpose.
}

\subsubsection{Donee Engine}
\begin{figure}[t!]
    \centering
    \includegraphics[width=0.75\columnwidth]{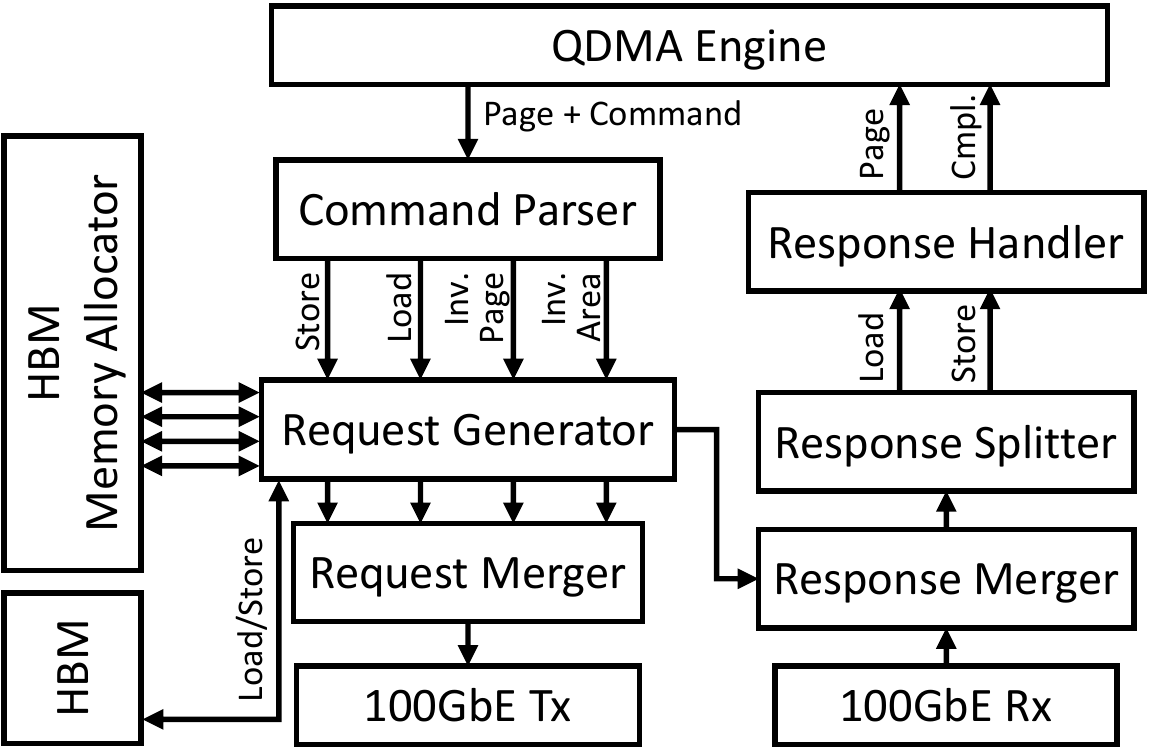}
    \\
    \caption{Block design of the donee engine}
    \label{fig:donee-engine-block-design}
\end{figure}

The block diagram of the donee engine is illustrated in Figure~\ref{fig:donee-engine-block-design}.

\noindent
\textbf{Command Parser:}
The command parser is responsible for interpreting commands and forwarding them to
the corresponding request generators. Each command has its own request generator.
To forward words to the correct request generator, the command parser reads the queue
ID (QID) field of the word, which is encoded by the kernel. As the command parser
is aware of the QIDs of each operation, it can forward words to the correct request
generator. Store commands require additional processing in the command parsing stage
not to mix up requests between queues. As a store command has multiple words to convey
a 4KB page, and the ordering between words is not guaranteed, words are stacked in
FIFO queues so that a command can be sent as a complete packet.

\noindent
\textbf{Request Generator:}
A request generator is responsible for parsing commands and taking required actions.
The store request generator stores pages in the local HBM or remote memory. The target
memory tier is determined by the kernel and encoded to the command field (\texttt{target\_tier}).
The memory allocation status is managed by the on-board HBM memory allocator, which
will be described in the following of this subsection. However, to speed up memory
allocation, the store request generator pre-allocates memory and caches the addresses
of several free blocks. When a store request to the local HBM fails, the request
is silently redirected to the donor engine so that memory allocation is done at the
donor node. If it fails to find free memory, then the donee stores the page to a
local swap device. The load request generator loads a page from the local HBM if
the page is at the local HBM. Otherwise, it will forward commands to the donor engine.
The invalidate page request generator and invalidate area request generator deallocate
pages from the local HBM and send commands to deallocate pages from remote memory.

\noindent
\textbf{On-board HBM Memory Allocator:}
The on-board HBM memory allocator is responsible for the allocation and deallocation
of on-board HBM memory. The memory allocation granularity is 4KB. The memory allocator
is implemented with a bitmap. As the size of HBM is 8GB, the size of the bitmap is
2,097,152-bit (256KB). To allow concurrent access to the table, \texttt{ARRAY\_PARTITION}
pragma has been applied to the table with a cyclic option and factor of 32.

\noindent
\textbf{Response Handler:}
The response handler reads returned responses and writes \texttt{completions} or
pages to the donee. Only two commands have responses: store and load. The completion
of page invalidation does not have to be identified. The identification of page invalidation
results in wasted hardware resource and network bandwidth. A store response is a
single-word response that contains the tier and address of the stored page. A load
packet has 65 words, where the first word has the metadata, and the following 64 words
have the page content. The writes from the store response handler and load response
handler are arbitrated in a round-robin manner.

\subsubsection{Donor Engine}
\begin{figure}[t!]
    \centering
    \includegraphics[width=0.85\columnwidth]{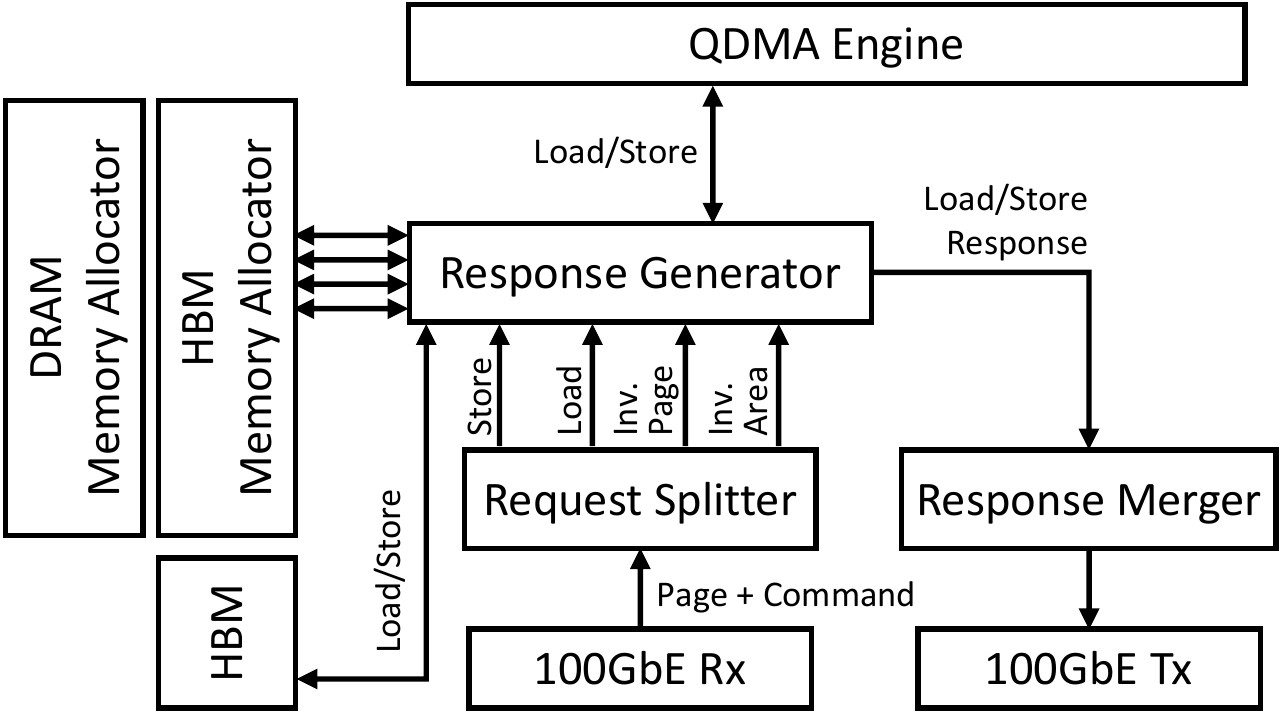}
    \caption{Block design of the donor engine}
    \label{fig:donor-block-design}
\end{figure}

Unlike the donee engine, where commands are sent from the host side, the donor engine
receives commands from the network interface. Figure~\ref{fig:donor-block-design}
presents the block design of the donor engine.

\noindent
\textbf{Request Splitter \& Response Generator:}
After receiving requests from the network module, the requests are split by the request
splitter. The request splitter reads the command field and forwards requests to the
corresponding response generators. The response generators are different from the
donee-side request generators in two aspects. First, the response generators coordinate
not only with the on-board HBM memory allocator but also with the host DRAM memory
allocator. Second, the invalidate page response generator and invalidate area response
generator do not generate response packets. The store response generator and load
response generator access donated memory, which is reserved by the kernel. The donor
reserves memory with the kernel boot parameter, \texttt{memmap}. The donor engine
knows the starting address and the size of the donated memory.

\begin{figure}[t!]
    \centering
    \includegraphics[width=0.83\columnwidth]{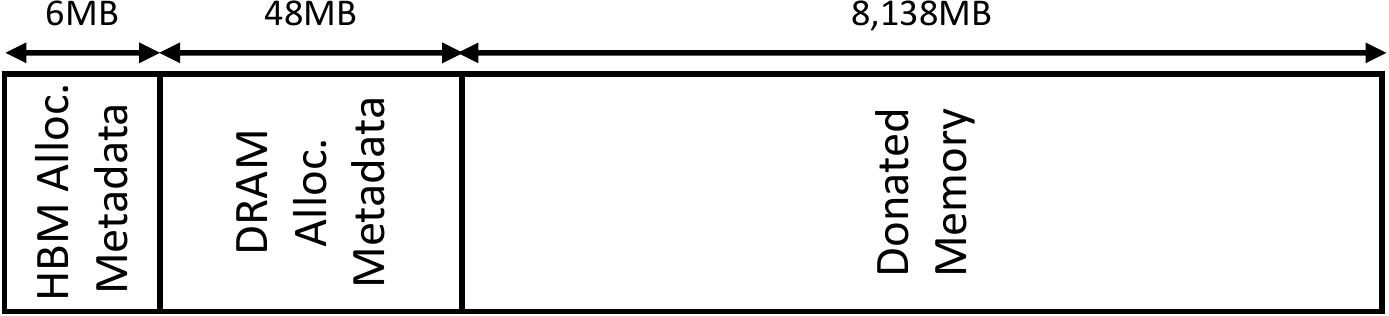}
    \caption{Donor-side HBM memory map}
    \label{fig:donor-hbm-memory-map}
\end{figure}

\noindent
\textbf{On-board HBM Allocator \& Host DRAM Allocator:}
The donor-side memory allocators have the same responsibility as the donee-side's,
which is to manage the allocation status of memory. However, there are two differences
from the donee-side memory allocator. First, the donor-side memory allocator has
to track the ownership of pages in addition to allocation status. The load response
generator looks up the memory allocator to confirm that the current load request
is trying to read a valid memory page that is owned by the requestor. The ownership
of pages is tracked with a machine ID (MID), whose size is 8-bit. As the metadata
size becomes eight times of the donee side's, it is not possible to hold all metadata
in the FPGA logic. Instead, the metadata is stored in the lower address of the HBM.
Figure~\ref{fig:donor-hbm-memory-map} presents the HBM memory map. Second, the donor
engine has the host DRAM memory allocator in addition to the on-board HBM memory
allocator.

\subsection{Software Implementation}
\label{sec:tdmem-software-implementation}
\subsubsection{Address Translation}
\begin{figure}[t!]
    \centering
    \includegraphics[width=\columnwidth]{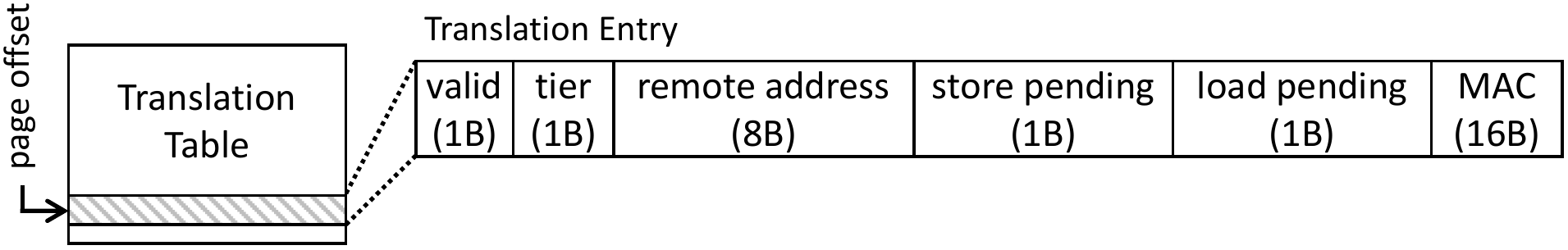}
    \\
    \caption{Translation table entry format}
    \label{fig:translation-table-format}
\end{figure}

\tdmem offloads the address translation responsibility to donees. Each donee has
a flat address translation table that maps a local address (swap offset) to a remote
physical. Figure~\ref{fig:translation-table-format} presents the format of the translation
entry. The \texttt{valid} field presents whether the entry has valid translation
information. A translation entry becomes valid when a store request completes. The
\texttt{tier} field presents the memory tier where the page is stored, and the \texttt{remote\_address}
field has the address of the page in the tier. The \texttt{store\_pending} and \texttt{load\_pending}
fields are used to coordinate with other concurrent swap requests. The fields are
used to prevent loading pages before store completion or invalidating pages before
page load completion. The \texttt{MAC} field is used for the authentication of decrypted
pages. The field is updated on encryption. The \texttt{MAC} field is compared with
the MAC that is generated on the decryption of the page.

\subsubsection{Software Page Encryption}
Pages are encrypted in the donee's kernel with AES-GCM on stores. The encryption
key and initialization vectors are generated in the initialization stage of the donee.
The AES-GCM algorithm takes associated data as an input for validation. The swap
offset of the page is used for the associated data. Pages are decrypted on page loads,
and the MAC is compared to the one generated during the encryption. MAC is stored
in the translation entry of the page, and if the MAC mismatches, \tdmem stops
serving the page fault, and the system halts.

\section{Evaluation}
\label{sec:tdmem-evaluation}
\subsection{Experimental Setup}
We evaluate the performance of \tdmem on a Linux system equipped with an FPGA card
and high-performance network card. The evaluation is conducted with two pairs of
machines. The first pair of machines is equipped with Mellanox ConnectX-5 for the
evaluation of fastswap. The second pair of machines has Xilinx Alveo U50 to evaluate
\tdmem. In each pair, one machine becomes a donor and the other becomes a donee.
Table~\ref{tab:tdmem-system-configurations} presents the machines and their configurations.
Table~\ref{tab:tdmem-workloads} presents the evaluated macrobenchmarks. The table
shows their memory footprint and the number of CPUs that they utilize.

\begin{figure}[t!]
    \centering
    \begin{minipage}{.45\columnwidth}
        \centering
        \includegraphics[width=\textwidth, height=3.33cm]{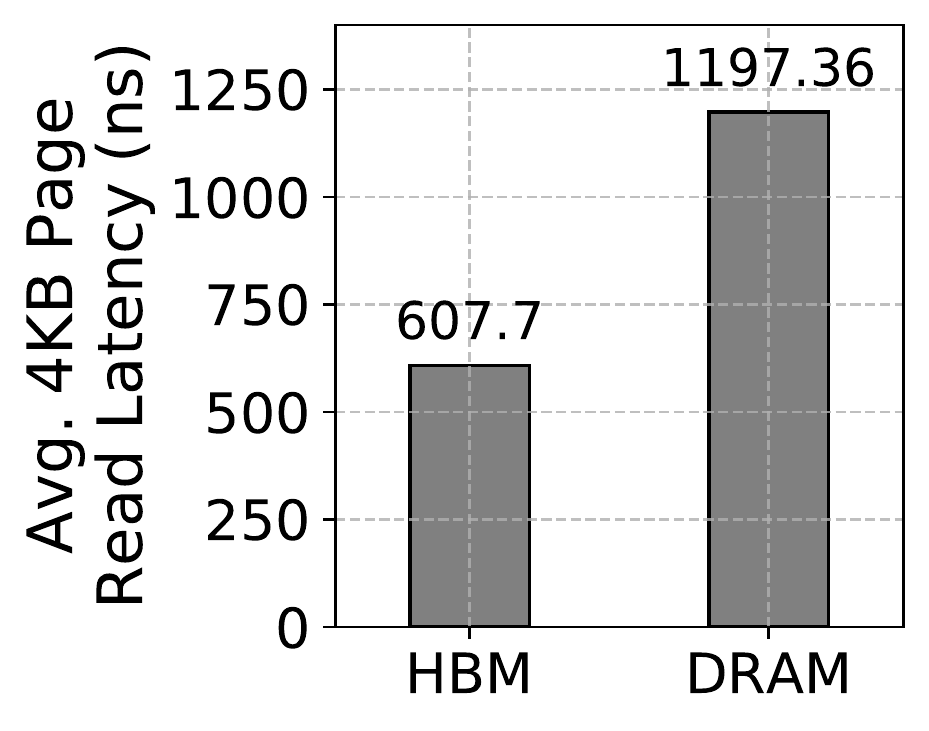}
        \\
        \caption{On-FPGA 4KB page read latency, which excludes the overhead of the software and network stack}
        \label{fig:on-fpga-page-read-latency}
    \end{minipage}
    \hfill
    \begin{minipage}{.45\columnwidth}
        \centering
        \includegraphics[width=\textwidth]{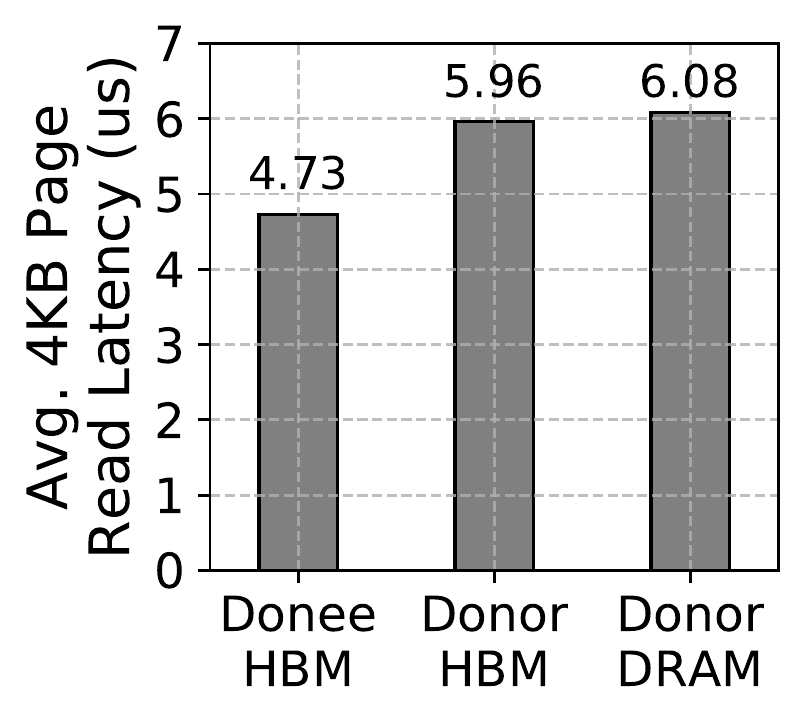}
        \\
        \caption{Donee-side 4KB page read latency, which includes the overhead of the software and network stack}
        \label{fig:donee-side-page-read-latency}
    \end{minipage}
\end{figure}

\newcolumntype{C}{>{\centering\arraybackslash}m{1.83cm}}
\newcolumntype{D}{>{\centering\arraybackslash}m{1.78cm}}

\begin{table}[t!]
    \footnotesize
    \centering

    \begin{tabular}{C|DDD}
        \toprule
         & Donor Host & Donee Host & Donee Guest \\
        \toprule
        Memory & DDR4 188GB & DDR4 125GB & DDR4 32GB \\
        Kernel & \multicolumn{2}{c}{5.9.0} & 5.10.0 \\
        Processors & \multicolumn{3}{c}{Intel(R) Xeon(R) CPU E5-2630 v4} \\
        OS & \multicolumn{3}{c}{Ubuntu 18.04} \\
        QDMA Driver & \multicolumn{3}{c}{2020.2} \\
        \bottomrule
    \end{tabular}

    \vspace{-0.1em}
    \caption{System configurations}
    \label{tab:tdmem-system-configurations}
    \vspace{-0.5em}
\end{table}

\begin{table}[t!]
    \footnotesize
    \centering
    \begin{tabular}{c|R{2.1cm}|R{1.3cm}}
        \toprule
        \multicolumn{1}{c}{\textbf{Workload Name}}
            & \multicolumn{1}{c}{\textbf{Mem. Footrprint (GB)}}
            & \multicolumn{1}{c}{\textbf{Num of CPUs}}\\ \toprule
        \texttt{tensorflow-inception} & 1.5 & 2 \\
        \texttt{kmeans} & 5.3 & 8 \\
        \texttt{quicksort} & 8.6 & 1 \\
        \texttt{in-memory-analytics} & 7.6 & 8 \\
        \texttt{graph-analytics} & 10.3 & 8 \\
        \texttt{xsbench} & 5.5 & 8 \\
        \bottomrule
    \end{tabular}
    \vspace{-0.1em}
    \caption{Macrobenchmarks and their memory footprint}
    \label{tab:tdmem-workloads}
\end{table}

\subsection{Microbenchmark Results}
\subsubsection{On-FPGA Page Read Latency}
Figure~\ref{fig:on-fpga-page-read-latency} presents the 4KB page read latency where
the software overhead is not included. The page read latency is measured by designing
and deploying a microbenchmark in the FPGA. The microbenchmark module reads 4KB pages
sequentially for a given range of memory addresses. The total elapsed cycles are
measured, and the elapsed cycles are divided by the number of pages read. As the
microbenchmark is designed to operate at 250MHz, the clock period is 4ns. The average
4KB page read latency is calculated by multiplying the clock period by the average
elapsed cycles. On average, HBM takes 607.7ns, and DRAM takes 1197.36ns to read a
4KB page.

\subsubsection{Donee-side Page Read Latency}
Unlike the previous experiment, which excludes the overhead of the software and network
stacks, this experiment includes them by measuring the page read latency in the donee-side
Linux kernel. Figure~\ref{fig:donee-side-page-read-latency} presents the average
page read latency in three types of memory tiers: donee-HBM, donor-HBM, and donor-DRAM.
The page read latency has been measured by sequentially reading 2,097,152 pages (8GB)
from the target memory tier. The pages are stored in the target memory before running
experiments, and the elapsed time has been measured with jiffies in the Linux kernel.
\texttt{CONFIG\_HZ}, which is the kernel configuration that defines the timer resolution,
is set to 250. The experiment result implies that the performance gap between HBM
and DRAM presented in Figure~\ref{fig:on-fpga-page-read-latency} is mostly hidden,
and the performance overhead comes from the network latency and page fault handling.

\begin{figure}[t!]
    \centering
    \begin{minipage}{.45\columnwidth}
        \centering
        \includegraphics[width=\textwidth]{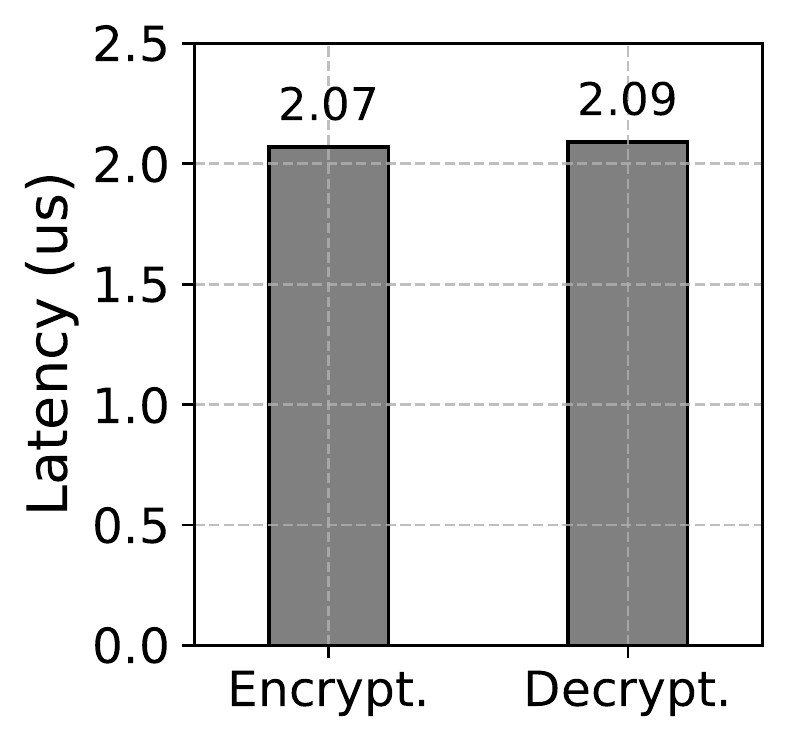}
        \\
        \caption{4KB page encryption latency and decryption latency\newline}
        \label{fig:page-encryption-latency-and-decryption-latency}
    \end{minipage}
    \hfill
    \begin{minipage}{.45\columnwidth}
        \centering
        \includegraphics[width=\textwidth]{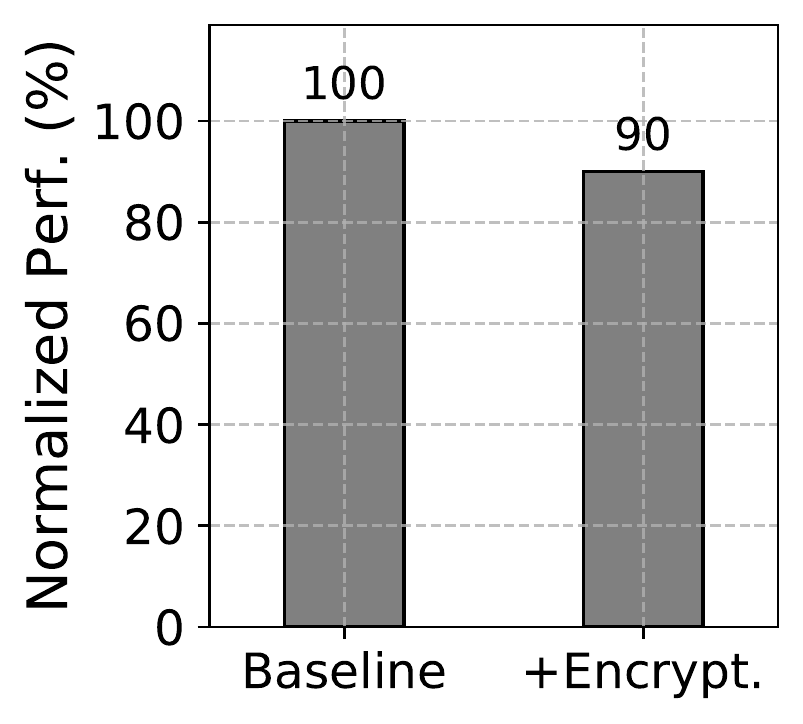}
        \\
        \caption{The impact of memory encryption on the microbenchmark performance}
        \label{fig:memory-encryption-impact-on-microbench-performance}
    \end{minipage}
\end{figure}

{
\tolerance=1
\emergencystretch=\maxdimen
\hyphenpenalty=10000
\hbadness=10000

\subsubsection{Page Encryption Latency}
\label{sec:evaluation-page-encryption-latency}
We measured the page encryption latency in the Linux kernel. The vanilla Linux kernel
has the \texttt{tcrypt} module, which evaluates the performance of encryption algorithms.
We evaluated the performance of \texttt{gcm(aes)} with the key size of 128-bit. Figure~\ref{fig:page-encryption-latency-and-decryption-latency}
presents the average page encryption and decryption latencies. It takes 2.07us for
4KB page encryption and 2.09us for 4KB page decryption. Although the latency seems
relatively high considering that the latency of page read is between 4-6us, most
of the latency can be hidden by the readahead mechanism of the Linux kernel.

Figure~\ref{fig:memory-encryption-impact-on-microbench-performance} shows the normalized
performance of a microbenchmark that loads 262,144 pages (1GB) sequentially from
the donee HBM. The normalized performance is defined as the performance normalized
to the baseline without encryption. The performance is the reverse of the elapsed
time. The performance of the microbenchmark with page encryption shows 90\% of the
performance without encryption. Most pages are readahead and decrypted in the swap
cache, effectively hiding the memory decryption latency.

\subsubsection{Memory Bandwidth}
In this experiment, we measure the memory bandwidth of \tdmem and compare it with
fastswap's. The memory bandwidth is measured with the STREAM benchmark. STREAM allocates
4GB memory and runs several sub-benchmarks to measure the bandwidth. Among the sub-benchmarks,
we use the Triad to measure the average memory bandwidth. We measure the memory bandwidth
of five configurations: \texttt{fastswap}, \texttt{\tdmem-HBM-plain}, \texttt{\tdmem-HBM-crypt},
\texttt{\tdmem-DRAM-plain}, and \texttt{\tdmem-DRAM-crypt}. \texttt{fastswap} presents
the case where the swapped-out pages are stored in remote memory with RDMA. The configurations
starting with \texttt{\tdmem} is run with \tdmem. The \texttt{HBM} and \texttt{DRAM}
keywords show the target memory tier where the swapped-out pages are stored at. The
\texttt{crypt} keyword means that the encryption latency is added on page loads.
The experiments are run with various local memory ratios, which include 90\%,
70\%, and 50\%. The local memory ratio is defined as the memory size in the donee
DRAM divided by the maximum memory footprint size. The memory footprint has been
measured with the cgroup's \texttt{memory.usage\_in\_bytes}.

\begin{figure}[t!]
    \centering
    \includegraphics[width=\columnwidth]{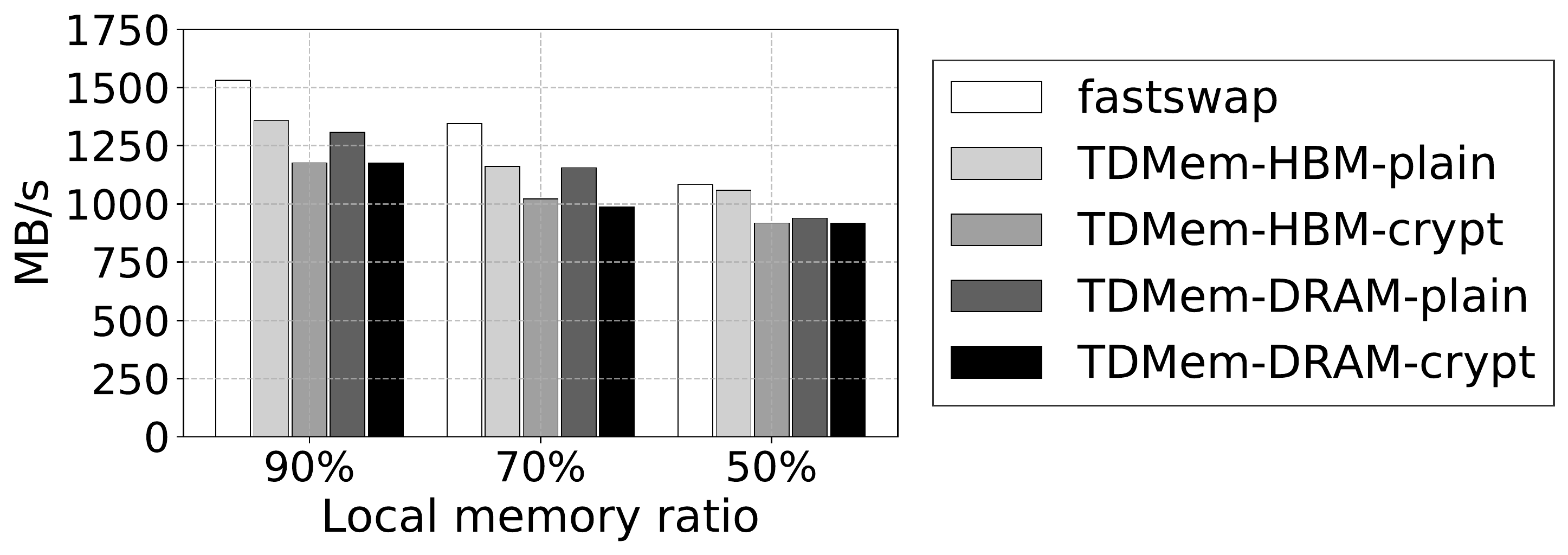}
    \\
    \caption{Memory bandwidth measured with the STREAM Triad benchmark}
    \label{fig:stream-memory-bandwidth}
\end{figure}

Figure~\ref{fig:stream-memory-bandwidth} shows the memory bandwidth with the configurations.
\texttt{fastswap} performs the best with the 90\% and 70\% local memory ratios. At
the 90\% local memory ratio, \texttt{fastswap}, \texttt{\tdmem-HBM-crypt}, and \texttt{\tdmem-DRAM-crypt}
present 1531.6MB/s, 1261.4MB/s, 1210.4MB/s, respectively. Please note that the memory
bandwidth gap between \texttt{\tdmem-HBM-crypt} and \texttt{\tdmem-DRAM-crypt} is
negligible, implying that the major bottleneck for the memory bandwidth is not the
memory itself. The memory bandwidth loss of \texttt{\tdmem-DRAM-crypt} compared to
\texttt{fastswap} are 21\%, 19\%, and 3\% at the 90\%, 70\%, and 50\% local memory
ratio, respectively.

\subsection{Macrobenchmark Results}
\textbf{Geomean of normalized performance:}
Figure~\ref{fig:geomean-normalized-performance} presents the geomean of normalized
performance of workloads for a given configuration. The normalized performance is
defined as the performance of a workload with a given configuration divided by the
performance of the workload run with 100\% local memory. The geomean of normalized
performance is the geomean of all normalized performance of workloads. Workloads
are run with five configurations: \texttt{fastswap}, \texttt{\tdmem-HBM-plain}, \texttt{\tdmem-HBM-crypt},
\texttt{\tdmem-DRAM-plain}, and \texttt{\tdmem-DRAM-crypt}. The experiments are run
while varying the local memory ratio between 40\% and 100\% with the 10\% step size.
Fastswap shows 64.5\% performance compared to the non-swapped run, and \tdmem experiences
negligible performance loss, presenting the 3-5\% lower performance compared to
fastswap. Please note that the performance overhead from software encryption is mostly
hidden because of the readahead mechanism in the Linux kernel, as we have shown in
Section~\ref{sec:evaluation-page-encryption-latency}.

\begin{figure}[t!]
    \centering
    \includegraphics[width=\columnwidth]{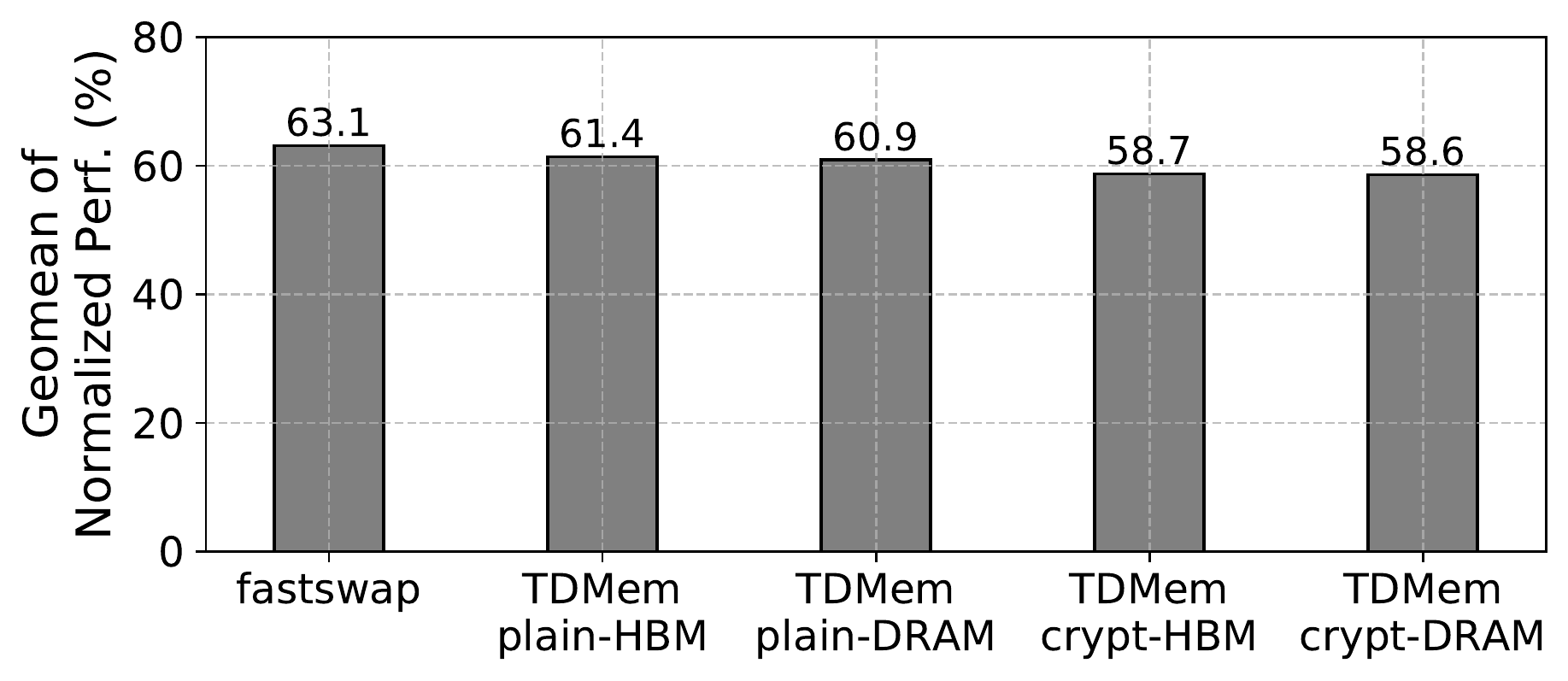}
    \\
    \caption{Geomean of normalized performance of workloads for each configuration}
    \label{fig:geomean-normalized-performance}
\end{figure}

\begin{figure}[t!]
    \centering
    \includegraphics[width=\columnwidth]{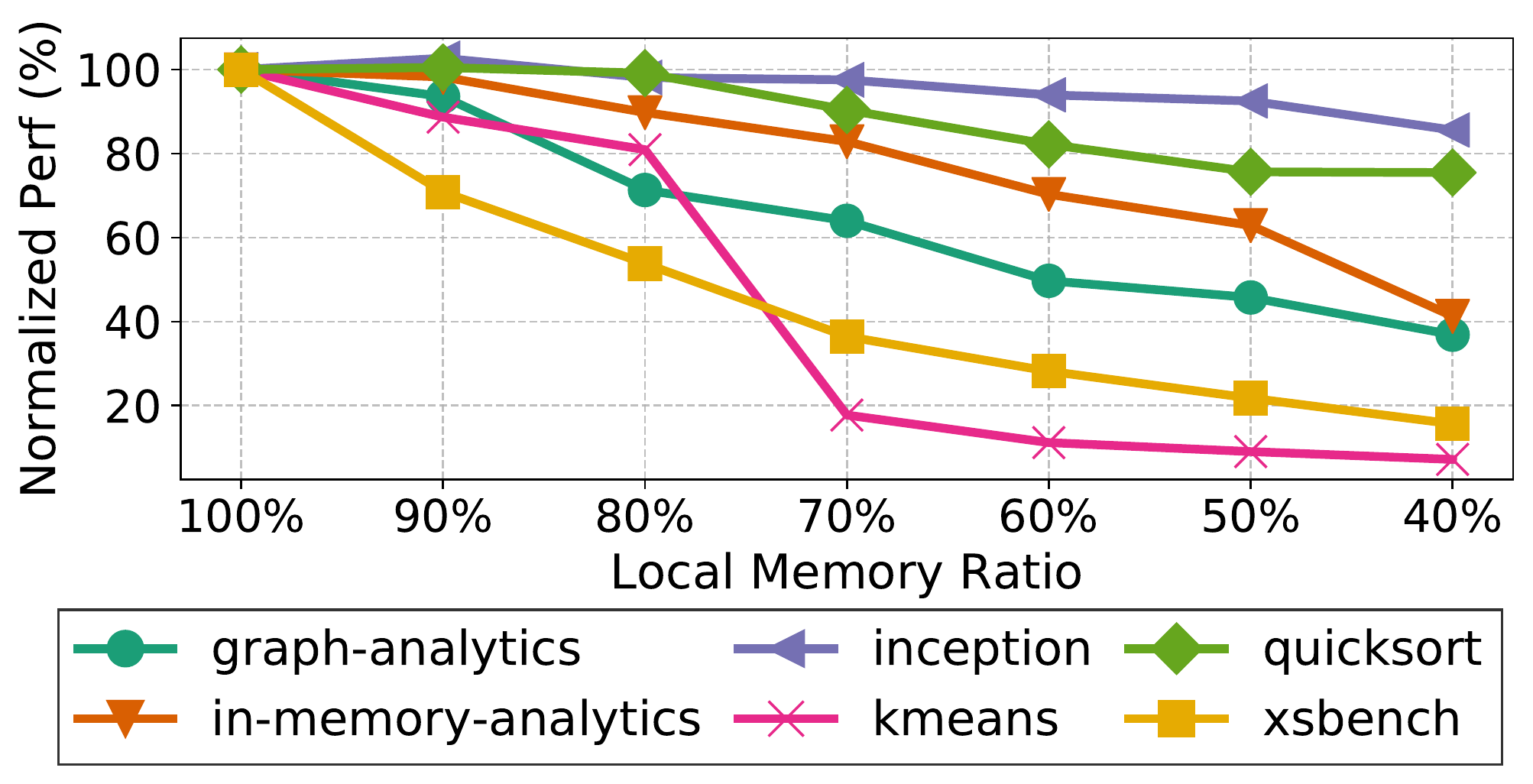}
    \\
    \caption{Normalized performance of workloads with various local memory ratios with \tdmem.
    Swapped out pages are stored in the donor DRAM}
    \label{fig:fpgaswap-normalized-performance}
\end{figure}

\noindent
\textbf{Normalized performance (\texttt{\tdmem-DRAM-crypt}):}
Among the experiment configurations presented in Figure~\ref{fig:geomean-normalized-performance},
we choose \texttt{\tdmem-DRAM-crypt} and illustrate workloads' performance degradation
while varying the local memory ratio in Figure~\ref{fig:fpgaswap-normalized-performance}.
The performance loss from losing local memory differs for each workload. While \texttt{in-memory-analytics}
presents 70\% of its performance with the 40\% local memory ratio, \texttt{kmeans}
shows 6.3\% of its performance. We qualitatively analyze the root cause with the
tools presented in prior studies~\cite{park2019profiling, 9252863}, which allow us
to analyze the memory access frequency. We found that the reason behind the different
sensitivity on the local memory ratio is the various memory access patterns and intensity.
While the memory footprint of \texttt{in-memory-analytics} is 10.6GB, only 2GBs of
memory is intensively accessed. On the other hand, the total allocated memory of
\texttt{kmeans} is intensively accessed, being more sensitive to the local memory
loss.

\begin{figure}[t!]
    \centering
    \includegraphics[width=\columnwidth]{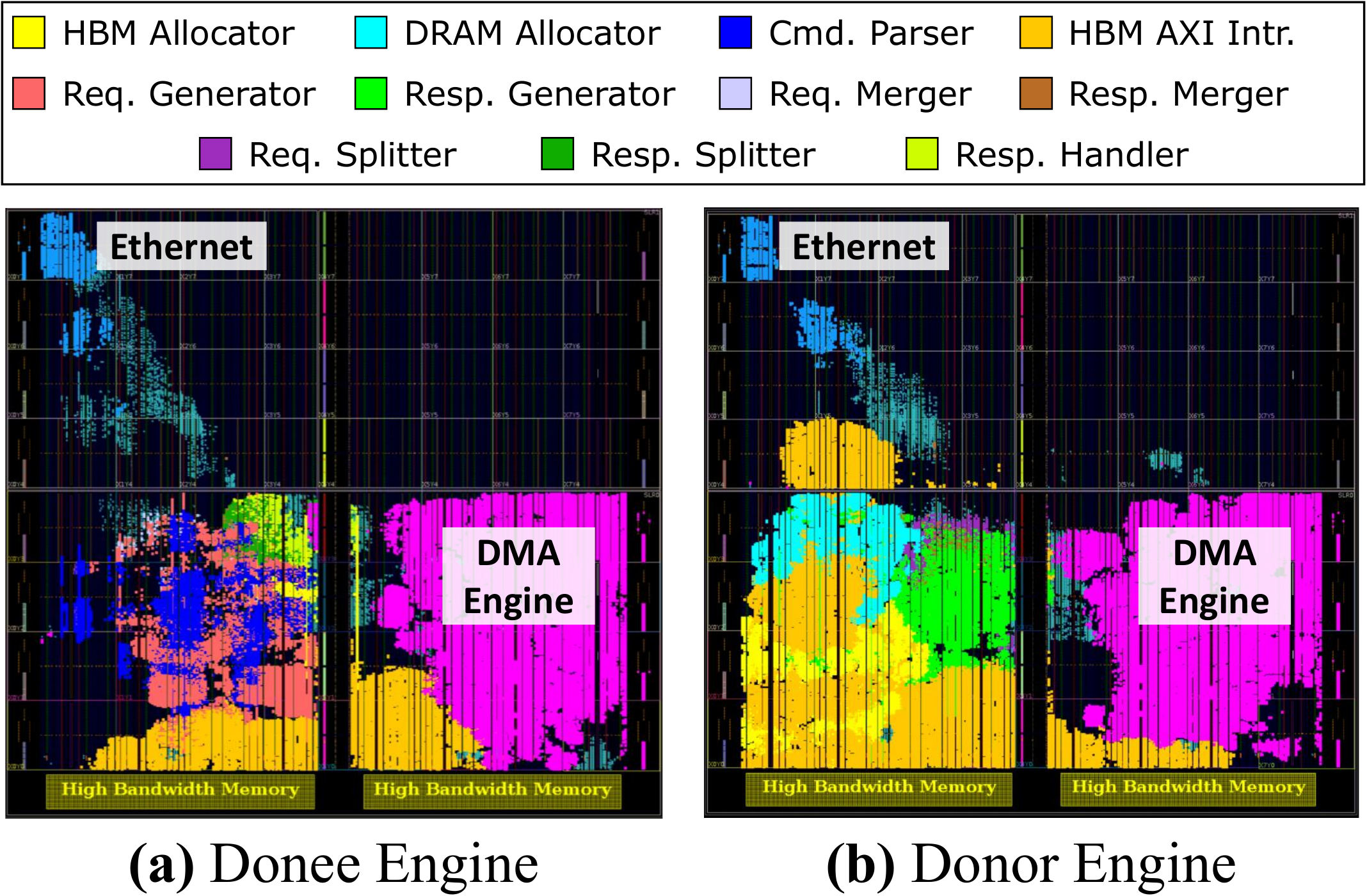}
    \\
    \caption{Floorplan of \tdmem}
    \label{fig:tdmem-floorplan}
\end{figure}

\begin{table}[t!]
    \footnotesize
    \centering
    \begin{tabular}{r|r|r|r|r|r|r|r|r|r|r}
        \toprule
        \multicolumn{11}{c}{\textbf{Donee Resource Utilization Breakdown (\%)}} \\ \toprule
        \multicolumn{1}{c}{\textbf{Component}} & \multicolumn{2}{c}{\textbf{LUT}} & \multicolumn{2}{c}{\textbf{LUTMEM}} & \multicolumn{2}{c}{\textbf{FF}} & \multicolumn{2}{c}{\textbf{BRAM}} & \multicolumn{2}{l}{\textbf{URAM}} \\ \toprule
        DMA Engine & \multicolumn{2}{r}{47.8} & \multicolumn{2}{r}{48.0} & \multicolumn{2}{r}{40.3} & \multicolumn{2}{r}{23.6} & \multicolumn{2}{r}{100.0} \\ \hline
        Cmd Parser & \multicolumn{2}{r}{6.7} & \multicolumn{2}{r}{0.0} & \multicolumn{2}{r}{11.8} & \multicolumn{2}{r}{35.5} & \multicolumn{2}{r}{0.0} \\ \hline
        Req Generator & \multicolumn{2}{r}{13.4} & \multicolumn{2}{r}{1.5} & \multicolumn{2}{r}{11.4} & \multicolumn{2}{r}{19.8} & \multicolumn{2}{r}{0.0} \\ \hline
        HBM Alloc & \multicolumn{2}{r}{0.4} & \multicolumn{2}{r}{0.0} & \multicolumn{2}{r}{0.3} & \multicolumn{2}{r}{15.8} & \multicolumn{2}{r}{0.0} \\ \hline
        HBM AXI Intr. & \multicolumn{2}{r}{10.3} & \multicolumn{2}{r}{37.7} & \multicolumn{2}{r}{19.5} & \multicolumn{2}{r}{0.0} & \multicolumn{2}{r}{0.0} \\ \hline
        Req Merger & \multicolumn{2}{r}{0.3} & \multicolumn{2}{r}{0.0} & \multicolumn{2}{r}{0.0} & \multicolumn{2}{r}{0.0} & \multicolumn{2}{r}{0.0} \\ \hline
        Ethernet & \multicolumn{2}{r}{0.8} & \multicolumn{2}{r}{1.5} & \multicolumn{2}{r}{1.4} & \multicolumn{2}{r}{0.0} & \multicolumn{2}{r}{0.0} \\ \hline
        Rsp Merger & \multicolumn{2}{r}{0.2} & \multicolumn{2}{r}{0.0} & \multicolumn{2}{r}{0.0} & \multicolumn{2}{r}{0.0} & \multicolumn{2}{r}{0.0} \\ \hline
        Rsp Spliter & \multicolumn{2}{r}{0.6} & \multicolumn{2}{r}{0.0} & \multicolumn{2}{r}{1.7} & \multicolumn{2}{r}{0.0} & \multicolumn{2}{r}{0.0} \\ \hline
        Rsp Handler & \multicolumn{2}{r}{0.6} & \multicolumn{2}{r}{0.0} & \multicolumn{2}{r}{1.7} & \multicolumn{2}{r}{0.0} & \multicolumn{2}{r}{0.0} \\ \hline
        Others & \multicolumn{2}{r}{18.8} & \multicolumn{2}{r}{11.3} & \multicolumn{2}{r}{11.9} & \multicolumn{2}{r}{5.3} & \multicolumn{2}{r}{0.0} \\
        \toprule
        \multicolumn{11}{c}{\textbf{Donor Resource Utilization Breakdown (\%)}} \\
        \toprule
        \multicolumn{1}{c}{\textbf{Component}} & \multicolumn{2}{c}{\textbf{LUT}} & \multicolumn{2}{c}{\textbf{LUTMEM}} & \multicolumn{2}{c}{\textbf{FF}} & \multicolumn{2}{c}{\textbf{BRAM}} & \multicolumn{2}{l}{\textbf{URAM}} \\ \toprule
        DMA Engine & \multicolumn{2}{r}{43.2} & \multicolumn{2}{r}{27.9} & \multicolumn{2}{r}{28.0} & \multicolumn{2}{r}{29.4} & \multicolumn{2}{r}{100.0} \\ \hline
        Req Splitter & \multicolumn{2}{r}{0.7} & \multicolumn{2}{r}{0.0} & \multicolumn{2}{r}{1.2} & \multicolumn{2}{r}{0.0} & \multicolumn{2}{r}{0.0} \\ \hline
        Rsp Generator & \multicolumn{2}{r}{10.1} & \multicolumn{2}{r}{1.7} & \multicolumn{2}{r}{9.9} & \multicolumn{2}{r}{51.4} & \multicolumn{2}{r}{0.0} \\ \hline
        DRAM Alloc & \multicolumn{2}{r}{6.4} & \multicolumn{2}{r}{2.4} & \multicolumn{2}{r}{7.7} & \multicolumn{2}{r}{3.6} & \multicolumn{2}{r}{0.0} \\ \hline
        HBM Alloc & \multicolumn{2}{r}{6.4} & \multicolumn{2}{r}{2.4} & \multicolumn{2}{r}{7.7} & \multicolumn{2}{r}{3.6} & \multicolumn{2}{r}{0.0} \\ \hline
        HBM AXI Intr. & \multicolumn{2}{r}{27.5} & \multicolumn{2}{r}{57.5} & \multicolumn{2}{r}{35.4} & \multicolumn{2}{r}{0.0} & \multicolumn{2}{r}{0.0} \\ \hline
        Rsp Merger & \multicolumn{2}{r}{0.2} & \multicolumn{2}{r}{0.0} & \multicolumn{2}{r}{0.0} & \multicolumn{2}{r}{0.0} & \multicolumn{2}{r}{0.0} \\ \hline
        Ethernet & \multicolumn{2}{r}{0.7} & \multicolumn{2}{r}{0.9} & \multicolumn{2}{r}{1.0} & \multicolumn{2}{r}{0.0} & \multicolumn{2}{r}{0.0} \\ \hline
        Others & \multicolumn{2}{r}{4.7} & \multicolumn{2}{r}{7.1} & \multicolumn{2}{r}{9.1} & \multicolumn{2}{r}{11.9} & \multicolumn{2}{r}{0.0} \\
        \bottomrule
    \end{tabular}
    \\
    \caption{Resource utilization breakdown of \tdmem}
    \label{tab:fpga-resource-utilization}
\end{table}

\subsection{FPGA Resource Utilization}
Figure~\ref{fig:tdmem-floorplan} illustrates the floorplan of the donee engine and donor
engine, and each component is filled with different colors. The donee engine consumes
18\% of LUT, 5\% of LUTRAM, 12\% of FF, 30\% of BRAM, and 2\% of URAM. The donor
engine accounts for 19\% of LUT, 9\% of LUTRAM, 17\% of FF, 24\% of BRAM, and 2\%
of URAM. As it can be seen in the figure, there is enough room for additional hardware
components. Table~\ref{tab:fpga-resource-utilization} presents the breakdown of logic
resource consumption for each engine. Each row presents the ratio of consumed logic
resources out of the total resource consumption of the engine. In both engines, DMA
engines are the major consumer of logic resources. In the donee engine, the HBM memory
allocator accounts for 15.8\% of the BRAM consumption of the donee engine. BRAM is
used to manage the memory allocation status of HBM memory. On the other hand, as
the donor engine manages the memory allocation status in the on-board HBM, the BRAM
usage of memory allocators is relatively low. The response generator consumes BRAM
for the free lists in the store response generator. All BRAMs in the response generator
are consumed by the store response generator.
}

\section{Related Work}
\label{sec:tdmem-related-work}
FPGA is drawing the attention of the computer architecture community and system community.
From the architecture perspective, FPGA is an attractive option to model custom accelerators
before diving into the ASIC design. Prototyping the performance of accelerators with
FPGAs has been done in neural network accelerators~\cite{sharma2016high, mahajan2016tabla,
hwang2020centaur, chang2021mix} and sparse matrix multiplication accelerators~\cite{elkurdi2008fpga,
lin2013design, grigoracs2016optimising, spaghetti2021}. From the system perspective,
it opens a new design space such as virtualization and resource management~\cite{zhang2017feniks,
khawaja2018sharing, shu2019direct, korolija2020abstractions}. Designing a system
with FPGAs is more than a pure academic study. Microsoft Catapult project~\cite{putnam2014reconfigurable,
caulfield2016cloud, fowers2018configurable, firestone2018azure} has shown that FPGAs
can improve the performance of real-world applications.

In swap-based disaggregated memory systems, the page replacement algorithm plays
an important role in identifying cold pages and swapping out them to remote memory.
Although there have been many theoretical studies to improve the page replacement
policies~\cite{robinson1990data, o1993lru, smaragdakis1999eelru, lee2001lrfu, megiddo2003arc,
bansal2004car, jiang2005clock, jiang2002lirs}, the Linux kernel page replacement
algorithm sticks to the LRU lists that mostly rely on heuristics in practice. Moreover,
the page replacement algorithm has been designed in the context of swapping pages
to a slow disk, which is quite different from the emerging memory systems. A few
recent studies tried to optimize the page replacement policy in the context of tiered
memory systems~\cite{ziyan2019nimble, 9252863}, and Google shared the improvement
for page replacement, the multi-generational LRU \cite{multi-generational-lru-lwn,
multi-generational-lru-patch} to the Linux kernel community.

\section{Conclusion}
\label{sec:tdmem-conclusion}
This paper proposes a new hardware-assisted memory disaggregation system, \tdmem.
It allows fine-grained page-level management of memory pools in donor nodes, while
access validation is enforced by the secure hardware engine protected from vulnerable
operating systems. In addition, it further secures the confidentiality of memory
pages with page address oblivious supports as well as encryption. The security features
of \tdmem cause a negligible performance overhead, which causes 4.4\% performance
degradation compared to the latest page-granular far memory system, fastswap. Although
our prototype is built upon an FPGA-based system, it can be designed for ASICs with
higher performance. Our evaluation with FPGA implementation showed that such fine-grained
secure disaggregated memory is feasible with comparable performance to the latest
software-based techniques.

\section{Acknowledgment}
This work was supported by the National Research Foundation of Korea (NRF-2019R1A2B5B01069816)
and the Institute for Information \& communications Technology Promotion (IITP2017-0-00466).
Both grants are funded by the Ministry of Science and ICT, Korea.

\bibliographystyle{plain}
\bibliography{references}

\end{document}